\documentclass[11pt]{article}

\usepackage[margin=1in]{geometry}

\usepackage{setspace}
\usepackage{times}

\usepackage{titlesec}
\titleformat*{\section}{\normalsize\bfseries}
\titleformat*{\subsection}{\normalsize\bfseries}
\titleformat*{\subsubsection}{\normalsize\bfseries}

\usepackage{graphicx}
\usepackage{amsmath, amsthm, amsfonts, amssymb}
\usepackage[hyphens]{url}
\usepackage[hidelinks]{hyperref}
\hypersetup{breaklinks=true}
\usepackage{cite}
\usepackage{xcolor}
\usepackage{bm}
\usepackage{booktabs}
\usepackage{subcaption}
\usepackage[font=small,labelfont=bf]{caption}
\usepackage[ruled,linesnumbered]{algorithm2e}
\newcommand{\eps}{\varepsilon}

\DeclareMathOperator*{\minimize}{minimize}

\newcommand\bbE{\ensuremath{\mathbb{E}}} 

\newcommand{\mat}[1]{\boldsymbol{#1}}
\renewcommand{\vec}[1]{\boldsymbol{\mathrm{#1}}}

\newcommand{\normof}[1]{\left\|#1\right\|}

\newcommand{\abs}[1]{\left\vert #1 \right\vert}

\newcommand{\RR}{\mathbb{R}}

\providecommand{\mD}{\ensuremath{\mat{D}}}

\providecommand{\ones}{\vec{e}}

\providecommand{\va}{\ensuremath{\vec{a}}}

\providecommand{\vecs}{\ensuremath{\vec{s}}}

\providecommand{\vu}{\ensuremath{\vec{u}}}

\providecommand{\vx}{\ensuremath{\vec{x}}}
\providecommand{\vy}{\ensuremath{\vec{y}}}

\newcommand{\States}{\mathcal{S}}
\newcommand{\Actions}{\mathcal{A}}
\newcommand{\Dist}{\mathcal{P}}

\newcommand{\StackelbergGame}{\mathcal{G}}
\newcommand{\Index}{\mathcal{I}}
\newcommand{\BR}{\mathrm{BR}}

\newcommand{\reg}{\mathrm{reg}}

\newcommand{\SE}{\mathrm{SE}}
\newcommand{\SSE}{\mathrm{SSE}}

\newcommand{\argmax}{\mathrm{argmax}}

\newcommand{\softmax}{\mathrm{softmax}}
\newcommand{\MF}{\mathrm{MF}}

\newcommand{\tol}{\texttt{tol}}

\newcommand{\MFStackelbergGame}{\StackelbergGame_\MF}

\theoremstyle{plain}
\newtheorem{theorem}{Theorem}[section]
\newtheorem{lemma}{Lemma}[section]
\newtheorem{corollary}[theorem]{Corollary}
\newtheorem{proposition}{Proposition}[section]

\theoremstyle{definition}
\newtheorem{definition}{Definition}[section]

\newtheorem{assumption}{Assumption}[section]

\theoremstyle{remark}
\newtheorem*{remark}{Remark}

\begin{document}

\renewcommand{\thefootnote}{\fnsymbol{footnote}}

\begin{center}
    \vspace*{-2em}
    \textbf{Learning in Stackelberg Markov Games}
    
    \vspace{1.5em}
    \textbf{Jun He}\footnote{Edwardson School of Industrial Engineering, Purdue University, West Lafayette. Email: \text{\href{mailto:he184@purdue.edu}{he184@purdue.edu}}}\hspace{2em}
    \textbf{Andrew L.~Liu}\footnote{Edwardson School of Industrial Engineering, Purdue University, West Lafayette. Email: \text{\href{mailto:andrewliu@purdue.edu}{andrewliu@purdue.edu}}}\hspace{2em}
    \textbf{Yihsu Chen}\footnote{Electrical and Computer Engineering, University of California, Santa Cruz. Email: \text{\href{mailto:yihsuchen@ucsc.edu}{yihsuchen@ucsc.edu}}}
\end{center}

\renewcommand{\thefootnote}{\arabic{footnote}}

\begin{abstract}
    Designing socially optimal policies in multi-agent environments is a fundamental challenge in both economics and artificial intelligence. This paper studies a general framework for learning Stackelberg equilibria in dynamic and uncertain environments, where a single leader interacts with a population of adaptive followers. Motivated by pressing real-world challenges such as equitable electricity tariff design for consumers with distributed energy resources (such as rooftop solar and energy storage), we formalize a class of Stackelberg Markov games and establish the existence and uniqueness of stationary Stackelberg equilibria under mild continuity and monotonicity conditions. We then extend the framework to incorporate a continuum of agents via mean-field (MF) approximation, yielding a tractable stationary Stackelberg-MF Equilibrium (SS-MFE) formulation. To address the computational intractability of exact best-response dynamics, we introduce a softmax-based approximation and rigorously bound its error relative to the true Stackelberg equilibrium. Our approach enables scalable and stable learning through policy iteration without requiring full knowledge of follower objectives. We validate the framework on an energy market simulation, where a public utility or a State utility commission sets time-varying rates for a heterogeneous population of prosumers. Our results demonstrate that learned policies can simultaneously achieve economic efficiency, equity across income groups, and stability in energy systems. This work demonstrates how game-theoretic learning frameworks can support data-driven policy design in large-scale strategic environments, with applications to real-world systems like energy markets.
\end{abstract}

\section{Introduction} \label{sec:intro}
Designing equitable and financially sustainable policies in rapidly decentralizing environments is a pressing challenge in mechanism design. A prominent real-world example is the design of electricity tariffs in modern power systems. As distributed energy resources (DERs), such as rooftop solar and home batteries, become increasingly common, standard rate structures are becoming severely distorted. Traditional residential rates typically combine a volumetric adder with a uniform fixed charge to recover a utility's transmission and distribution (T\&D) costs. However, as higher-income households adopt DERs and generate their own electricity, they contribute less to volumetric revenue. This forces the remaining conventional consumers to shoulder a larger share of grid costs, which drives rates higher and incentivizes even more DER adoption. The result is a financially unstable and socially regressive cycle where lower-income households, who depend most on grid electricity, bear a disproportionate share of the system's cost~\cite{chen2023optimaltariff}.

To address these inequities, regulators are exploring alternative rate structures, such as combining volumetric charges with income-graduated fixed charges. However, determining the optimal rate levels is highly complex. The regulator must anticipate how heterogeneous households will dynamically adapt their consumption and DER investments over time in response to the tariff, all under uncertainties in wholesale prices, renewable outputs, and demand. Fundamentally, this tariff design challenge is a prime example of a dynamic mechanism design problem that can be naturally modeled as a Stackelberg Markov game. In this framework, a leader (the regulator or utility) first commits to a strategy (the tariff), and followers (consumers and prosumers) rationally respond based on the leader’s choice.

Classical approaches to solving Stackelberg games rely on strong assumptions about agents' knowledge and rationality. They often require explicit, tractable models of the followers’ objectives and best-response behaviors, which are rarely available in realistic, stochastic environments like power markets. Recent advances in multi-agent reinforcement learning (MARL) have opened new possibilities for mechanism design in such settings. For instance, the AI Economist~\cite{zheng2022ai} demonstrates that a two-level reinforcement learning approach can help a planner and economic agents co-adapt, revealing emergent economic behaviors that are difficult to capture analytically.

Motivated by the rate design challenge, we abstract the problem and propose a general learning framework for Stackelberg Markov Games with infinite-horizon discounted rewards. We first establish the existence and uniqueness of a stationary Stackelberg equilibrium under mild regularity conditions in a two-agent setting. We then extend the framework to incorporate a single leader interacting with a continuum of competitive followers, modeled via a mean-field (MF) approximation. This seamlessly captures large-population environments—such as a grid of end-users—where each individual agent has negligible influence, and the leader seeks to shape collective behavior through policy design. To compute equilibria in both settings, we introduce a reinforcement learning algorithm that alternates between follower best-response learning and leader policy improvement, without requiring explicit knowledge of the followers’ reward functions.

The rest of the paper is organized as follows. Section~\ref{sec:related_work} reviews related work in both rate design and Stackelberg learning. Section~\ref{sec:single} formalizes the single-leader, single-follower Stackelberg Markov game and establishes equilibrium properties. Section~\ref{sec:learning} presents the learning algorithm and discusses its convergence. Section~\ref{sec:MFE} extends the framework to a mean-field population of followers. Finally, Section~\ref{sec:numerical} applies the framework back to the electricity tariff design problem using a standalone market in Oahu, Hawaii, which highlights its potential for real-world policy applications.
\section{Related Work} \label{sec:related_work}
Our work is first motivated by the pressing literature on equitable electricity rate design. First, the rapid adoption of DERs has highlighted the regressive nature of traditional volumetric recovery mechanisms, often referred to as the utility death spiral~\cite{chen2023optimaltariff}. To quantify and mitigate these inequities, recent literature has proposed metrics such as the energy expenditure incidence (EEI) which is the share of household income spent on electricity, and advocated for income-graduated fixed charges~\cite{chen21,chen2023optimaltariff,liu2025equitable}. While recent rate structures proposed by California's major utilities attempt to implement income-based fixed charges, the specific rate levels drew strong public criticism for being arbitrarily high~\cite{Conversation}. This underscores the critical need for the systematic, equilibrium-driven optimization approach we propose here. 

Methodologically, our work also bridges reinforcement learning (RL), game theory, and mechanism design, with the focus on Stackelberg games, which is a type of sequential games where a leader commits to a strategy and followers respond optimally. Classical solutions to these problems rely heavily on complete-information assumptions and bilevel optimization techniques~\cite{dempe2020bilevel}. To address stochastic and complex environments, recent works have turned to learning-based methods. The AI Economist~\cite{zheng2022ai}, for example, applies deep MARL to optimize tax policy in simulated economies, demonstrating the promise of data-driven mechanism design, though its focus is empirical and lacks theoretical guarantees. Complementing these empirical achievements, a parallel line of work has developed theoretical foundations for Stackelberg learning. \cite{bai2021sample} provide sample complexity bounds for learning Stackelberg equilibria under bandit feedback in general-sum games, while \cite{fiez2020convergence} analyze local convergence of gradient-based dynamics in continuous Stackelberg games. Furthermore, \cite{zhong2023can} study Stackelberg-Nash equilibria with myopic followers and propose provably efficient RL algorithms; however, their assumption of short-term follower behavior limits temporal expressiveness. In contrast, our framework provides global existence and uniqueness results and supports learning in infinite-horizon dynamic environments with fully strategic, forward-looking followers. By extending our model to mean-field populations, our work provides a theoretically grounded and scalable framework for learning Stackelberg equilibria in large-scale settings. Crucially, our approach relies only on observed policies and does not require explicit knowledge of agents’ reward functions, bridging the gap between rigorous game-theoretic foundations and practical policy design.
\section{The Single-Leader-Single-Follower Stackelberg Markov Game} \label{sec:single}
We now present the formal framework of Stackelberg Markov games and the corresponding learning algorithms. We begin with a single-leader, single-follower game in a dynamic and uncertain environment. This setting serves as the foundation for our theoretical results and for later extensions to large-population games.

A Stackelberg game is a sequential-move game in which one agent commits to a strategy first, anticipating its influence on the other’s response, and the second agent selects the best response after observing this commitment. In the classical formulation, such games are static and played under complete information, with equilibrium defined over a single strategic interaction. In contrast, we study Stackelberg interactions embedded in dynamic environments, formally, infinite-horizon discounted Markov games, where agents repeatedly interact with both one another and an evolving state governed by stochastic dynamics. While this setting resembles repeated games, we do not analyze the richer class of subgame perfect equilibria, which would require agents to condition strategies on full play histories and belief hierarchies. Instead, we adopt the perspective common in the literature on learning Nash equilibria in Markov games. That is, we focus on stationary strategies and aim to learn a static Stackelberg equilibrium, interpreted as the leader committing to a fixed policy and the follower adapting to it optimally within the environment. This formulation preserves the sequential structure of Stackelberg play while remaining tractable for reinforcement learning and policy optimization in a dynamic environment.  

We now formalize this setting. Let $\mathcal{I} = \{ L, F \}$ denote the two agents, where $L$ represents the first mover (leader) and $F$ the second mover (follower). For notational convenience, we write $-i$ to denote the opponent of agent $i$; that is, if $i = L$, then $-i = F$, and vice versa. Additionally, we use $\Dist(\mathcal{X})$ to denote the set of probability measures over a measurable set $\mathcal{X}$, and $x \sim \mathcal{Q}$ to indicate that $x$ follows distribution $\mathcal{Q}$. For sets $\mathcal{X}, \mathcal{Y}$, $\mathcal{X} \times \mathcal{Y}$ denotes their Cartesian product, and $\abs{\mathcal{X}}$ the cardinality of $\mathcal{X}$ if discrete. The definition of a Stackelberg Markov game is given below.

\begin{definition}[Single-Leader-Single-Follower Stackelberg Markov Game]
	A \emph{Stackelberg Markov game} with a single leader and a single follower is a tuple $\StackelbergGame := (\{ \States_i, \Actions_i, P_i, r_i, \gamma_i \}_{i \in \Index})$, where $\States_L$ and $\States_F$ are a (measurable) state spaces, and $\Actions_L$ and $\Actions_F$ are the action spaces of two agents. For each agent $i \in \Index$, the stochastic transition kernel $P_i: \States_i \times \Actions_L \times \Actions_F \to \Dist(\States_i)$ defines the probability distribution over next states, given current state $s_i$ and joint actions $(a_L, a_F)$. The reward functions $r_i : \States_L \times \States_F \times \Actions_L \times \Actions_F \to \RR$ specify agent $i$'s one-step payoff, and $\gamma_i \in [0,1)$ denotes its discount factor.
\end{definition}
In this paper, we focus on the case in which $\States_i$ and $\Actions_i$ are discrete and finite for each $i \in \Index$. The leader first observes its state $s_L \in \States_L$ and chooses an action $a_L \in \Actions_L$; then the follower observes its state $s_F \in \States_F$ and takes action $a_F \in \Actions_F$. The reward is then calculated through the reward functions $r_L$ and $r_F$. The leader and the follower take their actions according to their own policies $\pi_i \in \Pi_i := \{ \pi \mid \pi: \States_i \mapsto \Dist(\Actions_i)\}$. For simplicity, we define the joint state space $\States := \States_L \times \States_F$ and joint action space $\Actions := \Actions_L \times \Actions_F$. We let $\vecs = (s_L, s_F) \in \States$ and $\va = (a_L, a_F) \in \Actions$ denote the tuples of joint states and joint actions, respectively. The value function for each agent $i$ starting with initial joint states $\vecs$ is defined as follows:
\begin{equation} 
    V_i(\vecs; \pi_i, \pi_{-i}) := \bbE\Big[\sum_{t=0}^{\infty}\gamma_i^t r_i(\vecs_t, \va_t)  \Big\vert \vecs_0 = \vecs \Big], \label{eqn:value-lf}
\end{equation}
subject to $s_{i, t+1} \sim P_i(s_i, \va), a_{i, t} \sim \pi_i(s_{i, t})$, where the expectation is taken according to both agents' policies $\pi_L, \pi_F$, and the transition kernels $P_L, P_F$. Provided that the other player chooses $\pi_{-i}$, the goal for each player $i$ is to find the best policy $\pi_i^*$ that maximizes its value function with initial state $\vecs$ for all $\pi_i$:
\begin{equation} \label{eqn:value-best}
    V_i(\vecs; \pi_i^*, \pi_{-i}) \geq V_i(\vecs; \pi_i, \pi_{-i}).
\end{equation}
To facilitate the analysis of optimal policies as defined in \eqref{eqn:value-best}, and to ensure the existence of such solutions, we introduce the following assumption.
\begin{assumption}[Continuity and Boundedness] \label{assump:cts-bdd}
    Each agent $i$'s reward function $r_i(\vecs, \va)$ and the transition kernel $P_i(s_i, \va)$ are continuous in $\States_i$ and $\Actions$. In addition, the reward function is uniformly bounded; that is, there exists a finite number $R$ such that $\abs{r_i(\vecs, \va)} \leq R$, for all $\vecs \in \States$ and $\va \in \Actions$. 
\end{assumption}
This assumption is central to guaranteeing the existence of optimal stationary policies. Under standard conditions such as continuity, compactness, and boundedness, it is well established (e.g., \cite{puterman2014markov, agarwal2019reinforcement}) that stationary (i.e., time-invariant, memoryless) policies suffice for optimality in infinite-horizon discounted Markov games. We therefore focus exclusively on stationary policies, which simplifies the analysis and reflects both standard practice and real-world settings, where agents typically use fixed policies that depend only on the current state, not on time or history. 

In Stackelberg games, the leader has to anticipate (and respond to) the follower's best response. This requires us to define the follower's best response correspondence $\BR : \States \times \Pi_L \rightrightarrows \Pi_F$, which maps the joint state and the leader’s policy to the set of optimal followers' policies. That is, for any $\vecs$ and $\pi_L$,
\begin{equation}
    \BR(\vecs, \pi_L) := \argmax_{\pi_F \in \Pi_F} V_F(\vecs, \pi_F, \pi_L). \label{eqn:br-follower}
\end{equation}
Another useful notation is for leader's reduced value function, that is defined for any $\vecs$ and $\pi_L$:
\begin{equation}
    J_L(\vecs, \pi_L) = \max_{\pi_F\in\BR(\pi_L)} V_L(\vecs, \pi_L,\pi_F). \label{eqn:J-reduced-leader}
\end{equation}
For notation brevity, we omit the joint state arguments, and write $\BR(\pi_L)$ and $J_L(\pi_L)$. Specifically, a \emph{stationary Stackelberg equilibrium} (SSE) in the game $\StackelbergGame$ is defined as follows.
\begin{definition}[SSE]\label{def:se}
Given a Stackelberg Markov game $\StackelbergGame$, a pair of stationary policies $(\pi_L^{\SSE},\pi_F^{\SSE})$ forms an SSE, when the leader finds its optimal policy $\pi_L^\SSE \in \argmax_{\pi_L \in \Pi_L} J_L(\pi_L)$ and as a result, the follower's best response gives $\pi_F^\SSE \in \BR(\pi_L^\SSE)$.
\end{definition}

\subsection{Equilibrium Existence}
To establish the existence of an SSE, the following assumptions are needed:
\begin{assumption}[State, Action and Policy Spaces]\label{assump:spaces}
    For each agent $i$, both state and action spaces, $\States_i$ and $\Actions_i$, are finite. The stationary policy spaces $\Pi_i$ are endowed with the product topology induced by $\ell_1$ metric.
\end{assumption}
\begin{assumption}[Regular Best-Response]\label{assump:br}
    The follower's best-response correspondence $\BR$ has nonempty, compact and convex values and is upper hemicontinuous (UHC) in $\pi_L$.
\end{assumption}
The existence of an SSE can be formally stated in the following theorem:
\begin{theorem}[Existence of an SSE]\label{thm:sse-existence}
Under Assumptions~\ref{assump:cts-bdd},~\ref{assump:spaces} and~\ref{assump:br}, there exists at least one SSE.
\end{theorem}
To proceed with the proof of Theorem~\ref{thm:sse-existence}, the following lemmas are needed:
\begin{lemma}[Nonempty, Compact and Convex of Policy Spaces]\label{lemma:policy}
    Under Assumption~\ref{assump:spaces}, $\Pi_i$ is nonempty, compact and convex, for each $i \in \Index$.
\end{lemma}
\begin{proof}
Fix an agent $i\in\Index$. By definition, $\Pi_i := \{ \pi | \pi: \States_i \mapsto \Dist(\Actions_i)\} = \prod_{s_i\in\States_i} \Dist(\Actions_i)$ where $\Dist(\Actions_i)$ denotes the probability simplex over the finite action set $\Actions_i$. The product term is the $\abs{\States_i}$-fold Cartesian product that is commonly used as an alternative definition for sets of functions, as the product of copies of $\Dist(\Actions_i)$ by treating each state $s_i \in \States_i$ as an index (see Section 3 of Chapter 2 in \cite{hrbacek1999introduction}.) First, each simplex $\Dist(\Actions_i)$ contains at least the deterministic policy and hence is nonempty. $\Pi_i$, as the finite Cartesian product of nonempty sets, is therefore nonempty. Second, because $\Actions_i$ is finite, the simplex $\Dist(\Actions_i)$ is a closed and bounded subset of $\mathbb R^{|\Actions_i|}$, and therefore compact. As a result, $\Pi_i$, as the product of finitely many compact spaces, is compact. Finally, each simplex $\Dist(\Actions_i)$ is convex. The Cartesian product of convex sets is convex: if $\pi,\pi'\in\Pi_i$ and $t\in[0,1]$, then for each state $s$, $(t\pi + (1-t)\pi')(s)= t\pi(s) + (1-t)\pi'(s) $ lies in $\Dist(\Actions_i)$. Hence $t\pi+(1-t)\pi'\in\Pi_i$, which proves convexity.
\end{proof}

\begin{lemma}[Continuity of the value function]
\label{lem:V-continuous}
Under Assumptions~\ref{assump:cts-bdd} and~\ref{assump:spaces}, for each agent $i$ and each state $s_i$, the value function $V_i$ as in~\eqref{eqn:value-lf} is a continuous function of $(\pi_L,\pi_F)\in\Pi_L\times\Pi_F$.
\end{lemma}
\begin{proof}
Fix agent $i$. First we note that the policy vector $\pi_i = [\pi_i(a_i | s_i)]_{s_i \in \States_i, a_i \in \Actions_i} \in \RR^{\abs{\States_i}\abs{\Actions_i}}$ due to the finiteness of $\States_i$ and $\Actions_i$. We write the expected reward and transition kernel under $(\pi_L,\pi_F)$ as follows:
\begin{align*}
    & \overline r_i(\vecs, \pi_i, \pi_{-i}) = \sum_{\va} \pi_L(a_L\mid s_L) \pi_F(a_F\mid s_F) r_i(\vecs, \va), \\
    & \overline P(\vecs' \mid \vecs, \pi_L,\pi_F) = \sum_{\va} \pi_L(a_L\mid s_L) \pi_F(a_F\mid s_F) \cdot P_L(s_L' \mid s_L, \va) P_F(s_F' \mid s_F, \va).
\end{align*}
Since $\States_i$ and $\Actions_i$ are finite, both $\overline r_i$ and $\overline P$ are linear function and hence continuous on $(\pi_L,\pi_F)$. By stacking values over all states, the Bellman equation in vector form can be written as follow:
\begin{equation*}
    V_i(\pi_L,\pi_F) = \overline r_i(\pi_L,\pi_F) + \gamma \overline P(\pi_L,\pi_F) V_i(\pi_L,\pi_F),
\end{equation*}
which can be equivalently written as $(I - \gamma \overline P(\pi_L,\pi_F)) V_i(\pi_L,\pi_F) = \overline r_i(\pi_L,\pi_F)$, where $I$ is the identity matrix. Because $\overline P(\pi_L,\pi_F)$ is a stochastic matrix, its operator norm
satisfies $\normof{\overline P}_\infty = 1$, and since $\gamma\in[0,1)$ the matrix $I-\gamma\overline P(\pi_L,\pi_F)$ is invertible. Indeed one can verify this using the Neumann series representation $(I-\gamma\overline P)^{-1} = \sum_{k=0}^\infty \gamma^k \overline P^k$ which converges when the operator norm is less than 1, and $\normof{(I-\gamma\overline P)^{-1}} \leq \sum_{k=0}^\infty \gamma^k = \frac{1}{1 - \gamma}.$ Now for any policy pairs $(\pi_L,\pi_F)$ and $(\pi_L',\pi_F')$, with corresponding $\overline r,\overline P,V$ and $\overline r',\overline P',V'$. From the linear representation we have $V - V' = (I-\gamma\overline P)^{-1}\big(\overline r - \overline r' + \gamma(\overline P' - \overline P)V'\big)$. Taking the infinity norm and using the operator bound gives:
\begin{equation*}
    \normof{V - V'}_\infty \leq \frac{\normof{\overline r - \overline r'}_\infty + \gamma \normof{\overline P' - \overline P}_\infty \normof{V'}_\infty}{1 - \gamma} \leq \frac{\normof{\overline r - \overline r'}_\infty}{1-\gamma} +  \frac{\gamma R \normof{\overline P - \overline P'}_\infty}{(1-\gamma)^2},
\end{equation*}
where the second inequality uses the fact that rewards are bounded by a constant $R$, with which we have $\normof{V'}_\infty \leq R/(1-\gamma)$. Finally, due to the continuity of $\overline r$ and $\overline P$, the right-hand side goes to 0 whenever $(\pi_L',\pi_F')\to(\pi_L,\pi_F)$. Therefore the value function $V_i$ is continuous on $\Pi_L\times\Pi_F$.
\end{proof}
\begin{lemma}[Maximum Theorem's Lemma~\cite{Aliprantis2006_17}]\label{lemma:max-thm}
    Let $\phi: \mathcal X \rightrightarrows \mathcal Y$ be an UHC correspondence between topological spaces with nonempty compact values, and let $f: \text{Gr}(\phi) \to \RR$ be upper semicontinuous (USC), where $\text{Gr}(\phi) = \{(x, y) \in \mathcal X \times \mathcal Y | y \in \phi(x) \}$. Define the function $m: \mathcal X \to \RR$ by $m(x) = \max_{y \in \phi(x)} f(x, y)$. Then the function $m$ is USC.
\end{lemma}
\begin{theorem}[Generalized Extreme Value Theorem ~\cite{Aliprantis2006_2}]\label{thm:extreme-value}
    A real-valued USC function on a compact set attains a maximum value, and the nonempty of set of maximizers is compact.
\end{theorem}
Now we are ready to prove the existence of SSE.
\begin{proof}[Proof of Theorem~\ref{thm:sse-existence}]
Fix the joint state $(s_L,s_F)$. For simplicity, we drop the states arguments in this proof. Under Assumptions~\ref{assump:cts-bdd}, \ref{assump:spaces} and \ref{assump:br} the follower best-response correspondence $\BR$ is nonempty, compact-valued and UHC. Moreover the leader value function $V_L$ is continuous  and hence USC in $(\pi_L, \pi_F)$ from Lemma~\ref{lem:V-continuous}. By Lemma~\ref{lemma:max-thm}, the leader's reduced value function $J_L(\cdot)$ as defined in~\eqref{eqn:J-reduced-leader} is upper semi-continuous (USC) on the compact set $\Pi_L$, and the set of maximizers $\argmax_{\pi_F \in \BR(\pi_L)} V_L(\pi_L, \pi_F)$ is non-empty and compact-valued on $\Pi_L$. By Theorem~\ref{thm:extreme-value}, the real-valued WSC function $J_L$ attains its maximum on the compact space $\Pi_L$. Thus there exists $\pi_L^*\in\Pi_L$ such that, $J_L(\pi_L^*) = \max_{\pi_L\in\Pi_L} J_L(\pi_L)$. By the definition of $J_L$, there exists at least one follower policy $\pi_F^* \in \BR(\pi_L^*)$ achieving the inner maximum. Therefore, $\pi_L^*$ maximizes the leader's payoff anticipating the follower's best-response set, and $\pi_F^* \in \BR(\pi_L^*)$, which is the definition of an SSE.
\end{proof}
\section{Learning in Stackelberg Games} \label{sec:learning}

We now turn to the algorithmic problem of learning an SSE in dynamic environments. We consider a learning setting in which the interaction induces a two-time-scale separation: the follower reacts on a fast time scale, while the leader updates more slowly based on the follower's observed behavior. Our goal in this section is to formalize a general RL framework with this two-time-scale structure and to establish conditions under which the resulting learning framework converges to an SSE.

\subsection{Two-Time-Scale RL Framework} \label{sec:rl-framework}
We consider a single-leader single-follower Stackelberg game with policy-gradient-based RL. The leader and follower policies are parameterized by $\theta_L \in \Theta_L \subset \mathbb{R}^{n_L}$ and $\theta_F \in \Theta_F \subset \mathbb{R}^{n_F}$, respectively, where $n_L, n_F > 0$. The leader’s and follower’s policies are now notated as $\pi_{\theta_L}$ and $\pi_{\theta_F}$, respectively. Throughout this section, we slightly abuse notation by identifying the policy with its parameter and write $\pi_L \equiv \pi_{\theta_L}$ and $\pi_F \equiv \pi_{\theta_F}$ when no confusion arises. Accordingly, gradients and value functions with respect to $\theta_L$ and $\theta_F$ are understood as policy gradients induced by the corresponding parameterized policies. We consider a two-time-scale learning procedure, in which at each iteration $t$, the leader updates its policy with step-size $\alpha_t$ and the follower updates with step-size $\beta_t$:
\begin{align}
    \theta_{L,t+1} = \theta_{L,t} + \alpha_t \nabla_{\theta_L} V_L(\theta_{L,t},\theta_{F,t}), \label{eqn:leader-update} \\
    \theta_{F,t+1} = \theta_{F,t} + \beta_t \nabla_{\theta_F} V_F(\theta_{L,t},\theta_{F,t}). \label{eqn:follower-update}
\end{align}
The learning in~\eqref{eqn:leader-update} and \eqref{eqn:follower-update} operates on two distinct time scales that reflect the Stackelberg structure of the game. The follower repeatedly adapts its policy in response to the leader’s current strategy, while the leader updates its policy based on the follower’s observed behavior. To ensure that the leader has a meaningful gradient, the follower must react sufficiently fast so as to remain close to its best-response to the leader’s current policy. Accordingly, the leader is updated on a slower time scale, so that from the follower’s perspective the leader’s policy appears ``almost static.'' This time-scale separation enables the follower dynamics to converge to a stationary response for each fixed leader policy, while allowing the leader to evolve according to the induced reduced objective. Such a hierarchy of time scales is standard in stochastic approximation and forms the basis for the convergence analysis that follows. We first state the condition for the step-sizes:
\begin{assumption}[Two-Time-Scale Step Sizes]\label{ass:stepsizes}
The step sizes $\{\alpha_t\}$ and $\{\beta_t\}$ are positive and satisfy the following conditions:
\begin{equation}
    \sum_{t=0}^{\infty} \alpha_t = \sum_{t=0}^{\infty} \beta_t = \infty, \; \sum_{t=0}^{\infty} \alpha_t^2 < \infty, \; \sum_{t=0}^{\infty} \beta_t^2 < \infty, \text{ and } \lim_{t \to \infty} \frac{\alpha_t}{\beta_t} = 0.
\end{equation}
\end{assumption}
We note that the last condition enforces a strict separation of time scales between leader and follower updates. For completeness, we present the pseudo-code in Algorithm~\ref{alg:rl-tts} below:
\begin{algorithm}
    \caption{Two-Time-Scale Policy Gradient for Single-Leader–Single-Follower Stackelberg Games}
    \label{alg:rl-tts}
    \KwIn{Initial parameters $\theta_{L,0}, \theta_{F,0}$, step-size sequences $\{\alpha_t\}, \{\beta_t\}$, tolerance $\tol$.}
    \For{Iteration $t = 0, 1, 2, \cdots$}{
        Sample trajectories using policies $\pi_{\theta_{L,t}}$ and $\pi_{\theta_{F,t}}$ \;
        Compute stochastic policy gradient estimate $g_{F,t} = \nabla_{\theta_F} V_F(\theta_{L,t},\theta_{F,t})$ \;
        Update follower parameter: $\theta_{F,t+1} = \theta_{F,t} + \beta_t g_{F,t}$ \;
        Compute stochastic policy gradient estimate $g_{L,t} = \nabla_{\theta_L} V_L(\theta_{L,t},\theta_{F,t})$ \;
        Update leader parameter: $\theta_{L,t+1} = \theta_{L,t} + \alpha_t g_{L,t}$ \;
        If $\normof{\theta_{L,t+1} - \theta_{L,t}} + \normof{\theta_{F,t+1} - \theta_{F,t}} \le \tol$, exit the loop.
    }
    \Return $(\theta_L^t, \theta_F^t)$ as the learned SSE parameters.
\end{algorithm}

\subsection{Algorithm Convergence}\label{sec:convergence}
We now establish the convergence of the two-time-scale learning dynamics defined in~\eqref{eqn:leader-update} and~\eqref{eqn:follower-update}. The analysis follows the standard stochastic approximation framework with multiple time scales. The key idea is to separate the dynamics of the leader and the follower. On the fast time scale, the follower updates while the leader’s policy remains approximately constant. This allows the follower dynamics to be characterized by a limiting ordinary differential equation (ODE) parameterized by the leader’s policy. On the slow time scale, the leader evolves according to the follower's value function by composing its reward with the follower’s asymptotic best response. We proceed by first characterizing the asymptotic behavior of the follower updates for a fixed leader policy, then showing that the leader iterates track the gradient ascent dynamics of the value function. The following assumptions are made:
\begin{assumption}[Smoothness and Bounded Gradients] \label{ass:smooth-ts}
The value functions $V_L,V_F$ are continuously differentiable. The gradients are bounded and Lipschitz. That is, there exist finite constants $G_L,G_F$ such that $\normof{\nabla_{\theta_L}V_L(\theta_L,\theta_F)} \le G_L$ and $\normof{\nabla_{\theta_F}V_F(\theta_L,\theta_F)} \le G_F$ for all $\theta_L \in \Theta_L$ and $\theta_F \in \Theta_F$. Moreover, there exist $d_L, d_F > 0$ such that for all $\theta_L, \theta_L', \theta_F, \theta_F'$:
\begin{align}
& \normof{\nabla_{\theta_L}V_L(\theta_L,\theta_F)-\nabla_{\theta_L}V_L(\theta_L',\theta_F')} \le d_L(\normof{\theta_L-\theta_L'} + \normof{\theta_F-\theta_F'}), \\
& \normof{\nabla_{\theta_F}V_F(\theta_L,\theta_F)-\nabla_{\theta_F}V_F(\theta_L',\theta_F')} \le d_F(\normof{\theta_L-\theta_L'} + \normof{\theta_F-\theta_F'}).
\end{align}
\end{assumption}

\begin{assumption}[Strong Concavity of Follower's Value Function]\label{ass:strong-concave-ts}
The follower's value function is $\rho$-strongly concave in $\theta_F$. That is, there exists $ \rho>0$ such that for every $\theta_L$ and all $\theta_F,\theta_F'$,
\begin{equation}
    \left\langle \nabla_{\theta_F}V_F(\theta_L,\theta_F)-\nabla_{\theta_F}V_F(\theta_L,\theta_F'),\ \theta_F-\theta_F'\right\rangle \le - \rho \normof{\theta_F-\theta_F'}^2.
\end{equation}
\end{assumption}
We note that Assumptions~\ref{ass:smooth-ts} and~\ref{ass:strong-concave-ts} are stronger than Assumption~\ref{assump:br}, which only guarantees that $\BR$ is nonempty, compact, convex, and UHC in $\pi_L$. While Assumption~\ref{assump:br} suffices for SSE existence, it is not sufficient for analyzing learning dynamics driven by gradient-based updates. The smoothness and Lipschitz conditions in Assumption~\ref{ass:smooth-ts} ensure stability of the stochastic approximation iterates and allow the discrete-time updates to be well-approximated by their limiting ODEs on each time scale. The strong concavity condition in Assumption~\ref{ass:strong-concave-ts} further guarantees that, for any fixed leader policy, the follower admits a unique best response and that the fast-time-scale dynamics converge globally and exponentially to this response. Together, these assumptions allow the follower's asymptotic behavior to be represented by a single-valued, Lipschitz-continuous best-response mapping, which in turn ensures that the leader’s slow-time-scale updates track a well-defined gradient ascent flow. This structure is essential for establishing convergence of the coupled two-time-scale learning process. We formalize with the following lemmas:
\begin{lemma}[Lipschitz Best Response]\label{lem:BR-Lip-ts}
    Under Assumptions~\ref{ass:smooth-ts} and~\ref{ass:strong-concave-ts}, the best response $\BR$ is Lipschitz. That is, there exists $d_B > 0$ such that $ \normof{\BR(\theta_L)-\BR(\theta_L')} \le d_B\normof{\theta_L-\theta_L'} $ for all $\theta_L,\theta_L'\in\Theta_L$.
\end{lemma}
\begin{proof}
    Fix $\theta_L,\theta_L'\in\Theta_L$. Let $\theta_F^* := \BR(\theta_L)\in\argmax_{\theta_F\in\Theta_F}V_F(\theta_L,\theta_F)$ and $\theta_F^{*'} := \BR(\theta_L')\in\argmax_{\theta_F\in\Theta_F}V_F(\theta_L',\theta_F)$.
    By optimality, we have the first-order conditions $\nabla_{\theta_F}V_F(\theta_L,\theta_F^*)=0$ and $\nabla_{\theta_F}V_F(\theta_L',\theta_F^{*'})=0$. Subtracting them gives $\nabla_{\theta_F}V_F(\theta_L,\theta_F^*)-\nabla_{\theta_F}V_F(\theta_L,\theta_F^{*'}) =
    \nabla_{\theta_F}V_F(\theta_L',\theta_F^{*'})-\nabla_{\theta_F}V_F(\theta_L,\theta_F^{*'})$. We now take the inner product with $\theta_F^*-\theta_F^{*'}$ and use Assumption~\ref{ass:strong-concave-ts} (with $\theta_F=\theta_F^*$ and $\theta_F'=\theta_F^{*'}$, holding $\theta_L$ fixed) to obtain
    \begin{align*}
    \rho\normof{\theta_F^*-\theta_F^{*'}}^2 & \le \left\langle \nabla_{\theta_F}V_F(\theta_L,\theta_F^{*'})-\nabla_{\theta_F}V_F(\theta_L',\theta_F^{*'}), \theta_F^*-\theta_F^{*'}
    \right\rangle \\
    & \le \normof{\nabla_{\theta_F}V_F(\theta_L,\theta_F^{*'})-\nabla_{\theta_F}V_F(\theta_L',\theta_F^{*'})} \normof{\theta_F^*-\theta_F^{*'}}.
    \end{align*}
    If $\theta_F^*=\theta_F^{*'}$, we are done. Otherwise, we cancel $\normof{\theta_F^*-\theta_F^{*'}}$ to get
    \begin{equation*}
    \normof{\theta_F^*-\theta_F^{*'}} \le \frac{1}{\rho} \normof{\nabla_{\theta_F}V_F(\theta_L,\theta_F^{*'}) - \nabla_{\theta_F}V_F(\theta_L',\theta_F^{*'})} \le \frac{d_F}{\rho}\normof{\theta_L-\theta_L'}.
    \end{equation*}
    Thus $\BR$ is Lipschitz with Lipschitz constant $d_B:=d_F/\rho$.
\end{proof}

\begin{lemma}[Unique Best Response]\label{lem:uniqueBR-ts}
Under Assumption~\ref{ass:strong-concave-ts}, for each $\theta_L$, $\BR(\theta_L)$ exists and is unique.
\end{lemma}
\begin{proof}
    Existence follows from continuity of $V_F(\theta_L,\cdot)$ and compactness of $\Theta_F$. Strong concavity implies strict concavity, hence uniqueness.
\end{proof}

We additionally need the following assumptions:
\begin{assumption}[Bounded Policy Gradient Iterates]\label{ass:bounded-pg-iter}
    The iterates of the policy gradient are bounded: $\sup_{t}\left(\normof{\theta_{L,t}} + \normof{\theta_{F,t}}\right) < \infty$.
\end{assumption}
\begin{assumption}[Leader's Value Function]\label{ass:leader-value}
    The leader's value function $J_L$ is twice continuously differentiable in $\theta_L$, and the stationary points are non-degenerate. That is, for any stationary points $\theta_L^*$, it satisfies that $\nabla_{\theta_L} J_L(\theta_L^*) = 0$ and $\nabla_{\theta_L}^2 J_L(\theta_L^*)$ is invertible.
\end{assumption}
Assumption~\ref{ass:leader-value} ensures that the leader's ODE has well-behaved equilibrium structure. An immediate lemma can be established as follows:
\begin{lemma}[Leader's Internally Chain Transitive Invariant Sets]\label{lem:leader-ode}
    Under Assumption~\ref{ass:leader-value}, the only internally chain transitive invariant sets of the ODE $\dot{\theta}_L = \nabla_{\theta_L} J_L(\theta_L)$ are isolated equilibrium points, and each such equilibrium is isolated in the sense that there exists $\eps>0$ for which $\theta_L^*$ is the unique stationary point in the open ball $B_\eps(\theta_L^*) :=\{\theta_L:\normof{\theta_L-\theta_L^*}<\eps\}$.
\end{lemma}
\begin{proof}
    For simplicity, we drop the subscript $L$ for $\theta_L$. With slight abuse of notations (for this proof only), we define the gradient $g$ and Hessian $H$ as $g(\theta) := \nabla_{\theta} J_L(\theta)$ and $H(\theta) := \nabla^2_{\theta} J_L(\theta)$. At any stationary point $\theta^*$, by assumption, $g(\theta^*)=0$ and $H(\theta^*)$ is invertible. By inverse function theorem, there exists a neighborhood $B^{\theta^*}$ containing $\theta^*$ and a neighborhood $B^g$ containing $g(\theta^*)$ such that $g: B^{\theta^*} \mapsto B^g$ has a continuous inverse $g^{-1}: B^g \mapsto B^{\theta^*}$ which is differentiable for all points in $B^g$. In addition, $g$ is a bijective one-to-one mapping on $B^{\theta^*}$. Now, suppose $\theta \in B^{\theta^*}$ and $g(\theta)=0$. Then $g(\theta)=g(\theta^*)$. Since $g$ is one-to-one, we must have $\theta = \theta^*$. This implies $\theta^*$ is the unique zero of the gradient $g$ on the neighborhood $B^{\theta^*}$, which is an isolated stationary point of $J_L$. Finally, the equilibria of the ODE $\dot{\theta} = \nabla_\theta J_L(\theta) = g(\theta)$ are zeros of the gradient $g$, which forms a set of isolated stationary points of $J_L$.
\end{proof}
Finally, we can establish the main convergence theorem for the two-time-scale learning approach.
\begin{theorem}[Convergence of Two-Time-Scale Policy Gradient]\label{thm:converge-tts}
    Under Assumptions~\ref{assump:cts-bdd} -- \ref{ass:leader-value}, the two-time-scale policy gradient learning procedure in~\eqref{eqn:leader-update} and \eqref{eqn:follower-update} converges to the policy pair $(\theta_L^*, \theta_F^*)$, where $\theta_F^* = \BR(\theta_L^*)$ and $\theta_L^*$ is a stationary point of $V_L(\theta_L, \BR(\theta_L))$.
\end{theorem}
\begin{proof}
    The proof of Theorem~\ref{thm:converge-tts} follows the proof in \cite{Borkar2008_Ch6}. Specifically, under Assumptions~\ref{assump:cts-bdd} -- \ref{ass:leader-value}, one can first show that as $t \to \infty$, $(\theta_{L,t}, \theta_{F,t}) \to \{(\theta_L, \BR(\theta_L)):\theta_L \in \Theta_L\}$ (see Chapter 6, Lemma 1 of \cite{Borkar2008_Ch6}).That is, $\{\theta_{F,t}\}$ asymptotically track $\BR(\theta_{L,t})$. Then, under the same assumptions, one can show that $(\theta_{L,t}, \theta_{F,t}) \to (\theta_L^*, \BR(\theta_L^*))$ (see Chapter 6, Theorem 2 of \cite{Borkar2008_Ch6}) which tracks the leader's solution given by $\dot{\theta}_L = \nabla_{\theta_L} J_L(\theta_L)$. By Lemma~\ref{lem:leader-ode}, the $\theta_L^*$ is an isolated equilibrium point and hence by definition, a stationary point.
\end{proof}
\begin{corollary}[Convergence to SSE]
    Under Assumptions~\ref{assump:cts-bdd} -- \ref{ass:leader-value}, if the leader's value function $V_L$ is concave in $\theta_L$, the two-time-scale policy gradient converges to an SSE.
\end{corollary}
\begin{proof}
By Theorem~\ref{thm:converge-tts}, the two-time-scale policy gradient iterates converge to the set of stationary points. Under the concavity assumption, any stationary point of $V_L(\cdot,\BR(\cdot))$ is a global maximizer. Consequently, the limiting point $\theta_L^*$ maximizes the leader's objective given the follower’s best response, and the associated pair $(\theta_L^*, \BR(\theta_L^*))$ is an SSE. 
\end{proof}

\subsection{Reward Regularization} \label{sec:rew-regularization}
To establish convergence to a unique fixed point, one usually requires the follower's best-response operator $\argmax$ to be Lipschitz continuous and additionally, single-valued. To achieve this, we adopt the widely-used approach in RL: reward regularization, where it can accelerate convergence of policy gradient methods~\cite{cen2022fast} and enhance exploration and robustness~\cite{neu2017unified}. The regularized follower's reward function is defined as:
\begin{equation}
r^\reg_F(\vecs, \va) = r_F(\vecs, \va) + \alpha H(\pi_F(\cdot \mid s_F)), \label{eqn:reg-reward}
\end{equation}
where $\alpha > 0$ is a hyper-parameter, and $H(\cdot)$ is a $\rho$-strongly concave function where $\rho > 0$. A common choice is the Shannon entropy, given by $H(\pi_F(\cdot \mid s_F)) = -\sum_{a_F \in \Actions_F} \pi_F(a_F \mid s_F) \log \pi_F(a_F \mid s_F)$ for each $s_F$. We then analyze the game using the regularized value function for each joint state $\vecs$:
\begin{equation}
V^\reg_F(\vecs, \pi_F, \pi_L) := \bbE\Big[\sum_{t=0}^{\infty}\gamma_F^t r^\reg_F(\vecs_t, \va_t) \Big \vert  \vecs_0 = \vecs \Big]. \label{eqn:value-reg}
\end{equation}
Therefore, the follower now takes the best-response with the regularized value function:
\begin{equation}
    \BR^\reg(\vecs, \pi_L) := \argmax_{\pi_F \in \Pi_F} V^\reg_F(\vecs, \pi_F, \pi_L). \label{eqn:br-follower-reg}
\end{equation}
\begin{lemma}[Single-Valued Regularized Best Response] \label{lem:reg-unique}
    $\BR^\reg$ is single-valued.
\end{lemma}
\begin{proof}
    We first argue that the expected reward $\bbE[r(\vecs, \va)]$ is linear w.r.t. $\pi_F$. Indeed, the linearity is a direct consequence of the Lebesgue measure by viewing the distribution $\pi_F$ as the measure function. Then the sum of a linear function and a $\rho$-strongly concave function preserves the $\rho$-strong concavity of $V^\reg_F$ in $\pi_F$. Thus, $\BR^\reg$ is single-valued.
\end{proof}
Lemma~\ref{lem:reg-unique} shows that adding a strongly concave regularization term ensures the regularized value function $V^\reg_F$ is also strongly concave in $\pi_F$, which guarantees the uniqueness of the optimal solution $\pi_F^{\reg *} = \BR^\reg(\pi_L) = \argmax_{\pi_F} V^\reg_F(\pi_F, \pi_L)$. This implies that our two-time-scale learning approach converges to the point $(\pi_L^*, \BR^\reg(\pi_L^*))$. In the special case with Shannon entropy, the unique maximizer satisfies that for each $\pi_L$, the follower's policy obtained from $\BR^\reg$ has the following closed form:
\begin{equation}
    \pi_F^{\reg *}(a_F\mid s_F;\pi_L) =\frac{\exp\left(Q_F^{\pi_L}(s_F,a_F)/\alpha\right)}{\sum_{a_F'}\exp\left(Q_F^{\pi_L}(s_F,a_F')/\alpha\right)},
\end{equation}
which is known as the softmax function. Here, the follower's $Q$-function is defined as follows: for a pair of policy $(\pi_L, \pi_F)$,
\begin{equation}
Q_F^{\pi_L,\pi_F}(s_F,a_F) := \bbE\left[\sum_{t=0}^{\infty}\gamma_F^t r_F(\vecs_t,\va_t) \mid s_{F,0}=s_F, a_{F,0}=a_F \right].
\end{equation}
In the regularized case, when $\pi_F = \BR^\reg(\pi_L)$, we write $Q_F^{\pi_L} = Q_F^{\pi_L,\BR^\reg(\pi_L)}$. The optimal follower $Q$-function given $\pi_L$ is
\begin{equation}
Q_F^{\pi_L *}(s_F,a_F) := \max_{a_F\in\Actions_F} Q_F^{\pi_L}(s_F,a_F).
\end{equation}
While entropy regularization provides desirable properties such as robustness, smoothness, and improved exploration, it also introduces a systematic bias in the follower’s optimization problem. As a result, the convergent point of the two-time-scale learning dynamics corresponds to the regularized SSE rather than the true SSE of the original, unregularized game. In particular, the leader converges to an optimizer of the value function induced by the regularized best-response mapping $\BR^\reg$, instead of the exact best response. In the remainder of this section, we quantify the error gap between the regularized and the true SSE. We show that, under mild additional assumptions, this suboptimality gap can be explicitly bounded and made arbitrarily small by appropriately tuning the regularization parameter $\alpha$.
\begin{proposition}[Regularization Value Gap]\label{lem:entropy-gap}
When $H$ is the Shannon entropy, for every leader policy $\pi_L$ and initial state $\vecs$, it satisfies that $V_F^*(\vecs,\pi_L) - V_F(\vecs, \pi_F^{\reg *}(\pi_L),\pi_L) \leq \frac{\alpha\log|\Actions_F|}{1-\gamma_F}$, where $V_F^*$ denote the optimal value of the original value functions.
\end{proposition}
\begin{proof}
For any state $s_F$ and any distribution $\pi(\cdot\mid s_F)$, by Lemma~\ref{lem:shannon-bounds} (see Appendix), $0\le H(\pi(\cdot\mid s_F))\le \log\abs{\Actions_F}$.
Therefore for any $\pi_F$ and fixed $\pi_L$, by the definition of $V_F^\reg$:
\begin{equation*}
V_F(\pi_F,\pi_L) \le V_F^\reg(\pi_F,\pi_L) \le V_F(\pi_F,\pi_L)+\frac{\alpha\log|\Actions_F|}{1-\gamma_F}.
\end{equation*}
Let $\pi_F^*(\pi_L)\in\arg\max_{\pi_F}V_F(\pi_F,\pi_L)$.
By optimality of $\pi_F^{\reg *}(\pi_L)$ for the regularized objective,
\begin{equation*}
    V_F^{\reg *}(\pi_L) = V_F^\reg(\pi_F^{\reg *}(\pi_L),\pi_L) \ge V_F^\reg(\pi_F^*(\pi_L),\pi_L) \ge V_F(\pi_F^*(\pi_L),\pi_L)=V_F^*(\pi_L).
\end{equation*}
Using $V_F^\reg(\pi_F^{\reg *},\pi_L)\le V_F(\pi_F^{\reg *},\pi_L)+\frac{\alpha\log|\Actions_F|}{1-\gamma_F}$
gives the claimed bound $V_F^*(\pi_L)\le V_F(\pi_F^{\reg *}(\pi_L),\pi_L)+\frac{\alpha\log|\Actions_F|}{1-\gamma_F}$.
\end{proof}

\begin{assumption}[Lipschitz Leader's Value Function]\label{ass:leader-lip-follower}
There exists a constant $d_{V_L}>0$ such that for any leader policy $\pi_L, \pi_L'\in\Pi_L$, any two follower policies $\pi_F,\pi_F'\in\Pi_F$, and any initial state $\vecs$, the leader value function satisfies that $\abs{V_L(\vecs,\pi_L,\pi_F)-V_L(\vecs,\pi_L',\pi_F')} \leq d_{V_L} \left(\normof{\pi_L - \pi_L'} + \normof{\pi_F - \pi_F'}\right)$.
\end{assumption}

\begin{proposition}[Leader Value Gap]\label{prop:leader-gap-lip}
Define the secondary action set $\tilde{\Actions}_F := \{a_F \mid Q^{\pi_L}_F(a_F) \ne Q^{\pi_L *}_F \}$ as the set of non-optimal actions, and the action gap $\delta := \min_{a_F \in \tilde{\Actions}_F} \left(Q^{\pi_L *}_F - Q^{\pi_L}_F(a_F)\right)$ if $\abs{\tilde{\Actions}_F} > 0$, and $\delta := \infty$ otherwise. Let $\pi_L^*\in\argmax_{\pi_L}J(\pi_L)$ denote a (true) SSE optimal leader policy and $\pi_L^{\reg *}\in\argmax_{\pi_L}J^{\reg}_L(\pi_L)$ denote the leader policy optimal under the regularized response mapping. Under Assumptions~\ref{ass:leader-lip-follower}, for any initial state $\vecs$, the leader's value function satisfies that $\abs{J_L\left(\pi_L^*\right) - J^\reg_L\left(\pi_L^{\reg *})\right)} \le 2 \abs{\Actions_F} d_{V_L} e^{-\delta/\alpha}$.
\end{proposition}

\begin{proof}
Fix $\vecs$ and we drop this argument for notation brevity. By Assumption~\ref{ass:leader-lip-follower},
\begin{align*}
& \abs{J_L(\pi_L^*)-J^\reg_L(\pi_L^{\reg *})}  = \abs{\max_{\pi_L}J_L(\pi_L)-\max_{\pi_L}J_L^\reg(\pi_L)} \\ & \le \max_{\pi_L \in \Pi_L} \abs{J_L(\pi_L)-J^{\reg}_L(\pi_L)} = \max_{\pi_L \in \Pi_L} \abs{V_L\left(\pi_L,\pi_F^*(\pi_L)\right)-V_L\left(\pi_L,\pi_F^{\reg *}(\pi_L)\right)} \\
&\le d_{V_L} \max_{\pi_L \in \Pi_L} \normof{\pi_F^*(\pi_L)-\pi_F^{\reg *}(\pi_L)} \le 2 \abs{\Actions_F} d_{V_L} e^{-\delta/\alpha},
\end{align*}
where the last inequality follows Lemma~\ref{lemma:softmax-argmax-u} (see Appendix) by setting $c = 1/\alpha$. In addition, as $\alpha \to 0$, the gap $\abs{J_L(\pi_L^*)-J^\reg_L(\pi_L^{\reg *})} \to 0$.
\end{proof}
In summary, reward regularization guarantees convergence of the two-time-scale learning dynamics to a regularized SSE. We note that the regularization parameter $\alpha$ governs a trade-off between algorithmic stability and approximation accuracy: larger values of $\alpha$ improve smoothness and convergence properties, while smaller values recover the true SSE solution. In particular, as $\alpha \to 0$, the regularized SSE converges to the true SSE in value.

\section{Extension to Stackelberg Games with Mean-Field (MF) Followers}\label{sec:MFE}
We now consider the extension where there is one leader but an infinite number of followers in a (discounted) infinite-horizon Markovian setting. This formulation captures many real-world scenarios where a central authority (such as a platform, a regulator, or a policymaker) interacts with a large population of agents whose individual behaviors are negligible but whose aggregate effect shapes the system dynamics.  To formalize this setting, we adopt an MF approach in which followers are modeled as homogeneous and interchangeable. In the limit as the number of followers approaches infinity, each individual has vanishing influence on the aggregate behavior, which is captured by an MF distribution over states and actions. We assume the followers are competitive and analyze the interaction from the perspective of a single representative follower responding to the MF. This reduction preserves the coupling between individual incentives and population-level dynamics while simplifying the analysis. 

We also note that, the proposed framework and convergence analysis in this section are not restricted to homogeneous agents. In particular, the results extend directly to settings with heterogeneous populations that can be partitioned into a finite number of classes, where agents within each class are homogeneous but classes may differ in state spaces, reward functions, or dynamics \cite{mondal2022mfc}. This structure is common in practical applications and is precisely the setting considered in our numerical experiments. For such multi-class populations, the MF becomes a vector of class-specific distributions, and value functions, best responses, and MF update can be defined component-wise. The assumptions and proofs presented in this section apply independently to each class, and the overall system can be analyzed by stacking the class-level results into a joint vector form. For the ease of arguments, we drop the class indices, with the understanding that all assumptions and analysis work for each class.

For notational consistency, we retain the index set $\Index = \{L, F\}$, and denote the state and action spaces of the representative follower by $\States_F$ and $\Actions_F$, respectively. Let $\mu_{F,t} \in \mathcal{M}_F := \Dist(\States_F \times \Actions_F)$ denote an MF distribution at time $t$, representing the joint distribution of the population's states and actions in the infinite-agent limit for all $s \in \States_F, a \in \Actions_F$:
\begin{equation}
    \mu_{F,t}(s, a) := \lim_{N \to \infty} \frac{\sum_{j=1, j \ne i}^N \bm{1}_{(s_{F,t}^j, a_{F, t}^j)=(s, a)}}{N},
\end{equation}
where $N$ is the number of followers, and $(s_{F,t}^j, a_{F,t}^j)$ denotes the $j$-th follower's state and action pair. The indicator function $\bm{1}_{(s_{F,t}^j, a_{F,t}^j) = (s, a)} = 1$ if $(s_{F,t}^j, a_{F,t}^j) = (s, a)$, and 0 otherwise. The definition of a Stackelberg game with MF followers can be stated as follows.
\begin{definition}[Single-Leader-MF-Follower Stackelberg Markov Game]
     A Stackelberg Markov game with a single leader and a MF follower is a tuple $\MFStackelbergGame := (\{ \States_i, \Actions_i, P_i, r_i, \gamma_i \}_{i \in \Index})$, where $\States_L$ and $\States_F$ are a (measurable) state spaces, and $\Actions_L$ and $\Actions_F$ are the action spaces of two agents. For each agent $i \in \Index$, the stochastic transition kernel $P_i: \States_i \times \Actions \times \mathcal{M}_F \to \Dist(\States_i)$ defines the probability distribution over next states, given current state $s_i$, joint actions $(a_L, a_F)$ and follower's MF $\mu_F$. The reward functions $r_i : \States \times \Actions \times \mathcal{M}_F \to \RR$ specify agent $i$'s one-step payoff, and $\gamma_i \in [0,1)$ denotes its discount factor.
\end{definition}
The leader and the follower take their actions according to their own policies $\pi_i \in \Pi^{\MF}_i := \{ \pi \mid \pi : \States_i \times \mathcal{M}_F \to \Dist(\Actions_i) \}$. As the reward function and transition kernel are redefined with the MF as an additional argument, each agent's value function is also redefined as:
\begin{equation} 
    V_i(\vecs, \pi_i, \pi_{-i}, \mu_F) := \bbE\Big[\sum_{t=0}^{\infty}\gamma_i^t r_i(\vecs_t, \va_t, \mu_F) \Big\vert \vecs_0 = \vecs \Big], \label{eqn:value-mf}
\end{equation}
subject to $s_{i, t+1} \sim P_i(s_{i,t}, \va_t, \mu_F), a_{i, t} \sim \pi_i(s_{i, t}, \mu_F), \forall i \in \Index$, where the expectation is taken according to both agents' policies $\pi_L, \pi_F$, and the transition kernels $P_L, P_F$. The MF follower here assumes that the MF remains as $\mu_F$ throughout the entire lifetime. This can be achieved by maintaining a belief vector as in \cite{feng2025decentralized}, which can then be updated (adaptively) when a new leader policy is observed. Finally, the evolution of MF is a mapping $\Gamma: \mathcal{M}_F \times \Pi^\MF_L \times \Pi^\MF_F \to \mathcal{M}_F$ that maps from current MF and both agents' policies to the next MF, defined as follows for all $\mu_F \in \mathcal{M}_F, \pi_L \in \Pi^{\MF}_L, \pi_F \in \Pi^{\MF}_F$:
\begin{equation}\label{eq:MF_Update}
    \mu_F' := \Gamma(\mu_F, \pi_L, \pi_F),
\end{equation}
as a new component to the game. Then an equilibrium is defined as follows.
\begin{definition}[Stationary Stackelberg MF Equilibrium (SS-MFE)] \label{def:mfse}
    Given a Stackelberg Markov game $\MFStackelbergGame$ with MF followers, the tuple of the leader and the MF follower's stationary policies $(\pi_L^\SE, \pi_F^\SE, \mu_F^\SE)$ form an SS-MFE in $\MFStackelbergGame$, if the following conditions are satisfied for each state $s_L \in \States_L, s_F \in \States_F$, :
    \paragraph{(1) Follower:} For any policy $\pi_F \in \Pi^{\MF}_F$, $V_F(\vecs, \pi_F^\SE, \pi_L^\SE, \mu_F^\SE) \geq V_F(\vecs, \pi_F, \pi_L^\SE, \mu_F^\SE)$.
    \paragraph{(2) Consistency of Follower's MF:} The MF evolution satisfies that $\mu_F^\SE = \Gamma(\mu_F^\SE, \pi_L^\SE, \pi_F^\SE)$.
    \paragraph{(3) Leader:} For any policy $\pi_L \in \Pi^{\MF}_L$, $V_L(\vecs, \pi_L^\SE, \pi_F^\SE, \mu_F^\SE) \geq V_L(\vecs, \pi_L, \pi_F^\SE, \mu_F^\SE)$.
\end{definition}
The consistency condition in Item (2) indicates that when all followers adopt a policy in response to the assumed MF $\mu_F^\SE$, the resulting population distribution coincides with the assumed $\mu_F^\SE$. It is also useful to re-define the best response for follower as $\BR: \States \times \Pi^\MF_L \times \mathcal{M}_F \mapsto \Pi^\MF_F$:
\begin{equation}
    \BR(\vecs, \pi_L, \mu_F) := \argmax_{\pi_F \in \Pi_F} V_F(\vecs, \pi_F, \pi_L, \mu_F). \label{eqn:mf-br-follower}
\end{equation}
We simplify the notation by defining the map $\BR_{\mu}: \States \times \Pi^{\MF}_L \mapsto \Pi^{\MF}_F \times \mathcal{M}_F$, which is simply a composite update map from~\eqref{eqn:mf-br-follower} and~\eqref{eq:MF_Update} in the following form:
\begin{equation}
    \BR_{\mu}(\vecs, \pi_L) := \{ (\pi_F, \mu_F) \mid \pi_F \in \BR(\vecs, \pi_L, \mu_F), \mu_F = \Gamma(\mu_F, \pi_L, \pi_F) \}. \label{eqn:br-mu}
\end{equation}
The leader solves its value function, that is defined for any $\vecs$ and $\pi_L$:
\begin{equation}
    J_L(\vecs, \pi_L) = \max_{(\pi_F,\mu_F)\in\BR_\mu(\pi_L)} V_L(\vecs, \pi_L,\pi_F,\mu_F). \label{eqn:Jmf-reduced-leader}
\end{equation}

\subsection{Two-Time-Scale Policy Gradient}\label{sec:mf-tts}
We consider the same two-time-scale policy gradient learning approach as in Section~\ref{sec:learning}. We adopt the same parametrization, notation, and step sizes. The learning in~\eqref{eqn:leader-update} and~\eqref{eqn:follower-update} is now re-formulated as follows:
\begin{align}
    & \theta_{L,t+1} = \theta_{L,t} + \alpha_t \nabla_{\theta_L} V_L(\theta_{L,t},\theta_{F,t}, \mu_{F,t}), \label{eqn:mf-leader-update} \\
    & \theta_{F,t+1} = \theta_{F,t} + \beta_t \nabla_{\theta_F} V_F(\theta_{L,t},\theta_{F,t},\mu_{F,t}). \label{eqn:mf-follower-update} \\
    & \mu_{F,t+1} = \Gamma(\mu_{F,t}, \theta_{L,t+1}, \theta_{F, t+1}). \label{eqn:mf-update}
\end{align}
For completeness, we present the pseudo-code in Algorithm~\ref{alg:mf-rl-tts}.
\begin{algorithm}
    \caption{Two-Time-Scale Policy Gradient for Single-Leader–MF-Follower Stackelberg Games}
    \label{alg:mf-rl-tts}
    \KwIn{Initial parameters $\theta_{L,0}, \theta_{F,0}$, step-size sequences $\{\alpha_t\}, \{\beta_t\}$, tolerance $\tol$.}
    \For{Iteration $t = 0, 1, 2, \cdots$}{
        Sample trajectories using policies $\pi_{\theta_{L,t}}$ and $\pi_{\theta_{F,t}}$ \;
        Compute stochastic policy gradient estimate $g_{F,t} \approx \nabla_{\theta_F} V_F(\theta_{L,t},\theta_{F,t}, \mu_{F,t})$ \;
        Update follower parameter: $\theta_{F,t+1} = \theta_{F,t} + \beta_t g_{F,t}$ \;
        Compute stochastic policy gradient estimate $g_{L,t} \approx \nabla_{\theta_L} V_L(\theta_{L,t},\theta_{F,t}, \mu_{F,t})$ \;
        Update leader parameter: $\theta_{L,t+1} = \theta_{L,t} + \alpha_t g_{L,t}$ \;
        Update follower MF: $\mu_{F,t+1} = \Gamma(\mu_{F,t}, \theta_{L,t+1}, \theta_{F, t+1})$ \;
        If $\normof{\theta_{L,t+1} - \theta_{L,t}} + \normof{\theta_{F,t+1} - \theta_{F,t}} + \normof{\mu_{F,t+1} - \mu_{F,t}} \le \tol$, exit the loop.
    }
    \Return $(\theta_{L,t}, \theta_{F,t}, \mu_{F,t})$ as the learned SS-MFE parameters.
\end{algorithm}

To ensure algorithm convergence in the MF case, we adopt the following assumptions:
\begin{assumption}[Smoothness and Bounded Gradients] \label{ass:smooth-mf-ts}
The value functions $V_L,V_F$ are continuously differentiable. The gradients are bounded and Lipschitz. That is, there exist finite constants $G_L,G_F$ such that $\normof{\nabla_{\theta_L}V_L(\theta_L,\theta_F,\mu_F)} \le G_L$ and $\normof{\nabla_{\theta_F}V_F(\theta_L,\theta_F,\mu_F)} \le G_F$ for all $\theta_L \in \Theta_L$, $\theta_F \in \Theta_F$ and $\mu_F \in \mathcal{M}_F$. Moreover, there exist $d_L^\mu, d_F^\mu > 0$ such that for all $\theta_L, \theta_L', \theta_F, \theta_F', \mu_F, \mu_F'$:
\begin{align*}
& \normof{\nabla_{\theta_L}V_L(\theta_L,\theta_F,\mu_F)-\nabla_{\theta_L}V_L(\theta_L',\theta_F',\mu_F')} \le d_L^\mu(\normof{\theta_L-\theta_L'} + \normof{\theta_F-\theta_F'} + \normof{\mu_F-\mu_F'}), \\
& \normof{\nabla_{\theta_F}V_F(\theta_L,\theta_F,\mu_F)-\nabla_{\theta_F}V_F(\theta_L',\theta_F',\mu_F')} \le d_F^\mu(\normof{\theta_L-\theta_L'} + \normof{\theta_F-\theta_F'} + \normof{\mu_F-\mu_F'}).
\end{align*}
\end{assumption}
We also reuse Assumption~\ref{ass:strong-concave-ts} about follower's value function being $\rho$-strongly concave in $\theta_F$. Then, the following lemma holds:
\begin{lemma}[Lipschitz and Single-valued Best Response]\label{lem:BR-Lip-ts-mf}
    Under Assumptions~\ref{ass:smooth-ts} and~\ref{ass:strong-concave-ts}, the best response $\BR$ is Lipschitz and single-valued. That is, there exists $d_B^\mu > 0$ such that for all $\theta_L,\theta_L'\in\Theta_L$ and $\mu_F, \mu_F' \in \mathcal{M}_F$, we have $\normof{\BR(\theta_L, \mu_F)-\BR(\theta_L', \mu_F')} \leq d_B^\mu(\normof{\theta_L-\theta_L'} + \normof{\mu_F - \mu_F'})$.
\end{lemma}
\begin{proof}
    The result is a direct extension of Lemma~\ref{lem:BR-Lip-ts} where one can replace $\normof{\theta_L - \theta_L'}$ by $\normof{\theta_L - \theta_L'} + \normof{\mu_F - \mu_F'}$. As a result, the $\BR$ is Lipschitz with the constant $d_B^\mu:=d_F^\mu/\rho$. The single-valued-ness is also a direct consequence of strong concavity by Assumption~\ref{ass:strong-concave-ts}.
\end{proof}

\begin{assumption}[Lipschitz and Single-valued MF Update] \label{assump:mf-unique}
    For any $\mu_F \in \mathcal{M}_F, \pi_L \in \Pi^{\MF}_L, \pi_F \in \Pi^{\MF}_F$, the follower's MF update mapping $\Gamma(\mu_F, \pi_L, \pi_F)$ is Lipshcitz continuous and single-valued. That is, for any leader's policies $\pi_L, \pi_L'$, the follower's policies $\pi_F, \pi_F'$, and follower's MF $\mu_F, \mu_F'$, there exists a constant $d_\Gamma > 0$ such that $\normof{\Gamma(\mu_F, \pi_L, \pi_F) - \Gamma(\mu_F', \pi_L', \pi_F')} \leq d_\Gamma \left(\normof{\mu_F - \mu_F'}_1 + \normof{\pi_L - \pi_L'}_1 + \normof{\pi_F - \pi_F'}\right)$.
\end{assumption}
\begin{remark}[Interpretation and justification of Assumption~\ref{assump:mf-unique}]
In policy-gradient, policies are parameterized by finite-dimensional vectors, and the update is carried out in the parameter space rather than directly in policy space.
Assumption~\ref{assump:mf-unique} can then be equivalently stated with $\pi_L,\pi_F$ replaced by their parameters $\theta_L,\theta_F$. That is, $\normof{\Gamma(\mu_F, \theta_L, \theta_F) - \Gamma(\mu_F', \theta_L', \theta_F')} \leq d_\Gamma \left(\normof{\mu_F - \mu_F'}_1 + \normof{\theta_L - \theta_L'}_1 + \normof{\theta_F - \theta_F'}\right)$ for any $\theta_L, \theta_L', \theta_F, \theta_F', \mu_F, \mu_F'$. In addition, from a modeling perspective, assuming that the MF update mapping $\Gamma$ is single-valued reflects the fact that, given a current population distribution and fixed leader and follower policies, the next-step MF is uniquely determined by the underlying state-transition dynamics.
This is standard in MF control and MF game models, where the population evolution is governed by a deterministic flow induced by the policy and transition kernel. Similarly, the Lipschitz continuity of $\Gamma$ is a mild regularity condition expressing the stability of the population dynamics: small perturbations in the MF or in the agents’ policies lead to proportionally small changes in the next population distribution. 
\end{remark}

The remainder of the section focuses on proving the properties of the composite map $\BR_\mu$. If $\BR_\mu$ is Lipschitz and single-valued in $\theta_L$ as in the single-follower case's best response mapping described in Lemma~\ref{lem:BR-Lip-ts-mf} and~\ref{lem:uniqueBR-ts}, then all results from the single-leader-single-follower case follow.
\begin{assumption}[Lipschitz Constant Bound]\label{ass:lip-constant-mf}
    We assume $d_B^\mu + d_\Gamma \in (0, 1)$.
\end{assumption}
\begin{lemma}[Single-valued and Lipschitz Composite Map] \label{lem:composite}
    Suppose Assumptions~\ref{ass:strong-concave-ts}, \ref{ass:smooth-mf-ts}, \ref{assump:mf-unique} and \ref{ass:lip-constant-mf} hold. For every $\theta_L \in \Theta_L$, $\BR_\mu(\theta_L)$ defined in~\eqref{eqn:br-mu} is a singleton (\textit{i.e.} $\BR_\mu$ is single-valued), denoted by the pair $(\theta_F^*(\theta_L), \mu_F^*(\theta_L))$. Furthermore, the mapping $\BR_\mu$ is Lipschitz continuous. That is, for any $\theta_L, \theta_L'$, one has $\normof{\BR_\mu(\theta_L) - \BR_\mu(\theta_L')} \le d_\MF \normof{\theta_L - \theta_L'}$, where the Lipschitz constant is given by $d_\MF = \frac{d_B^\mu + d_\Gamma}{1 - (d_B^\mu + d_\Gamma)}$.
\end{lemma}
\begin{proof}
    Fix a state $\vecs$ and a leader policy $\theta_L$. For notational simplicity, we omit the $\vecs$ argument. Recall that by Definition~\eqref{eqn:br-mu}, a pair $(\theta_F, \mu_F)$ belongs to $\BR_\mu(\theta_L)$ if and only if it satisfies $\theta_F = \BR(\theta_L, \mu_F)$ and $\mu_F = \Gamma(\mu_F, \theta_L, \theta_F)$.
    Define the joint operator $\Psi_{\theta_L} : \Theta_F \times \mathcal{M}_F \to \Theta_F \times \mathcal{M}_F$ as:
    \begin{equation}
        \Psi_{\theta_L}(\theta_F, \mu_F) := \left( \BR(\theta_L, \mu_F), \Gamma(\mu_F, \theta_L, \theta_F) \right).
    \end{equation}
    We equip the product space $\Theta_F \times \mathcal{M}_F$ with the norm $\normof{(\theta_F, \mu_F)} = \normof{\theta_F} + \normof{\mu_F}_1$. Consider two pairs $z = (\theta_F, \mu_F)$ and $z' = (\theta_F', \mu_F')$. By the Lipschitz properties in Lemma~\ref{lem:BR-Lip-ts-mf} and Assumption~\ref{assump:mf-unique}:
    \begin{align*}
        &\normof{\Psi_{\theta_L}(z) - \Psi_{\theta_L}(z')} = \normof{\BR(\theta_L, \mu_F) - \BR(\theta_L, \mu_F')} + \normof{\Gamma(\mu_F, \theta_L, \theta_F) - \Gamma(\mu_F', \theta_L, \theta_F')} \\
        &\le d_B^\mu \normof{\mu_F - \mu_F'} + d_\Gamma \left(\normof{\mu_F - \mu_F'}_1 + \normof{\theta_F - \theta_F'}\right) = d_\Gamma \normof{\theta_F - \theta_F'} + (d_B^\mu + d_\Gamma) \normof{\mu_F - \mu_F'} \\
        &\le (d_B^\mu + d_\Gamma) \left( \normof{\theta_F - \theta_F'} + \normof{\mu_F - \mu_F'} \right) = (d_B^\mu + d_\Gamma) \normof{z - z'}.
    \end{align*}
    By the assumption that $d_B^\mu + d_\Gamma \in (0, 1)$, $\Psi_{\theta_L}$ is a contraction mapping. By the Banach Fixed-Point Theorem (see Theorem~\ref{thm:banach} in Appendix), there exists a unique fixed point $z^*(\theta_L) = (\theta_F^*(\theta_L), \mu_F^*(\theta_L))$ such that $z^* = \Psi_{\theta_L}(z^*)$, which implies $\BR_\mu(\theta_L)$ is single-valued. To show $\BR_\mu$ is Lipschitz in $\theta_L$, let $z^* = \Psi_{\theta_L}(z^*)$ and $z'^* = \Psi_{\theta_L'}(z'^*)$. Then:
    \begin{align*}
        & \normof{z^* - z'^*} = \normof{\Psi_{\theta_L}(z^*) - \Psi_{\theta_L'}(z'^*)} \le \normof{\Psi_{\theta_L}(z^*) - \Psi_{\theta_L}(z'^*)} + \normof{\Psi_{\theta_L}(z'^*) - \Psi_{\theta_L'}(z'^*)} \\
        &\le (d_B^\mu + d_\Gamma)\normof{z^* - z'^*} + \normof{\BR(\theta_L, \mu_F'^*) - \BR(\theta_L', \mu_F'^*)} + \normof{\Gamma(\mu_F'^*, \theta_L, \theta_F'^*) - \Gamma(\mu_F'^*, \theta_L', \theta_F'^*)}_1 \\
        &\le (d_B^\mu + d_\Gamma)\normof{z^* - z'^*} + (d_B^\mu + d_\Gamma) \normof{\theta_L - \theta_L'}.
    \end{align*}
    Rearranging the terms, we obtain $\normof{z^* - z'^*} \le \frac{d_B^\mu + d_\Gamma}{1 - (d_B^\mu + d_\Gamma)} \normof{\theta_L - \theta_L'}$, which proves that $\BR_\mu$ is Lipschitz with constant $d_\MF = \frac{d_B^\mu + d_\Gamma}{1 - (d_B^\mu + d_\Gamma)}$.
\end{proof}
We now present the convergence theorem for the single-leader-MF-follower Stackelberg Markov game.
\begin{theorem}[Convergence of Two-Time-Scale Policy Gradient]\label{thm:converge-mf-tts}
    Under Assumptions~\ref{assump:cts-bdd} -- \ref{ass:stepsizes}, \ref{ass:strong-concave-ts} -- \ref{ass:leader-value}, \ref{ass:smooth-mf-ts} -- \ref{ass:lip-constant-mf}, the two-time-scale policy gradient learning procedure in~\eqref{eqn:mf-leader-update} -- \eqref{eqn:mf-update} converges to $(\theta_L^*, \theta_F^*, \mu_F^*)$, where $(\theta_F^*, \mu_F^*) = \BR_\mu(\theta_L^*)$ and $\theta_L^*$ is a stationary point of $V_L(\theta_L, \BR_\mu(\theta_L))$. In addition, if $V_L$ is concave, the point is an SS-MFE.
\end{theorem}
\begin{proof}
    The proof follows the two-time-scale stochastic approximation framework for coupled systems \cite{Borkar2008_Ch6}. The learning dynamics can be viewed as three coupled processes. Given the step-size condition in Assumption~\ref{ass:stepsizes}, the leader's parameter moves on a slower time scale relative to the follower's parameter. We first consider the fast-scale dynamics. For a fixed $\theta_L$, the pair $(\theta_{F,t}, \mu_{F,t})$ is governed by the coupled updates \eqref{eqn:mf-follower-update} and \eqref{eqn:mf-update}. By Lemma~\ref{lem:composite}, we have established that the composite mapping $\BR_\mu(\theta_L)$ is single-valued and the joint operator is a contraction with constant $d_B^\mu + d_\Gamma < 1$. This implies that the fast-scale ODE 
    \begin{equation}
        \begin{pmatrix} \dot{\theta}_F \\ \dot{\mu}_F \end{pmatrix} = \begin{pmatrix} \nabla_{\theta_F} V_F(\theta_L, \theta_F, \mu_F) \\ \Gamma(\mu_F, \theta_L, \theta_F) - \mu_F \end{pmatrix}
    \end{equation}
    possesses a globally asymptotically stable equilibrium at $(\theta_F, \mu_F) = \BR_\mu(\theta_L)$. Consequently, as $t \to \infty$, the fast variables asymptotically track the set $\{(\theta_L, \theta_F, \mu_F) \mid (\theta_F, \mu_F) = \BR_\mu(\theta_L)\}$. Now, by substituting this stable equilibrium into the slow-scale update \eqref{eqn:mf-leader-update}, the leader's dynamics track the reduced ODE $\dot{\theta}_L = \nabla_{\theta_L} V_L(\theta_L, \BR_\mu(\theta_L))$, where $\BR_\mu$ is $d_\MF$-Lipschitz. Following the same logic as the proof of Theorem~\ref{thm:converge-tts}, the trajectory $\theta_{L,t}$ converges to a stationary point $\theta_L^*$ of the reduced leader's value function $J_L$. When $V_L$ is concave, $(\theta_L^*, \theta_F^*, \mu_F^*)$ satisfies all conditions of the SS-MFE in Definition~\ref{def:mfse}.
\end{proof}
\section{Numerical Experiments: Electricity Tariff Design} \label{sec:numerical}

To demonstrate the practical utility of our learning-based Stackelberg framework, we apply it to the problem of equitable electricity tariff design. We ground our experiments in a case study of Oahu, Hawaii, a standalone market with a high penetration of DERs and publicly available data, making it an ideal testbed for this environment. In our setup, the leader (a regulator or utility) determines a tariff structure comprising a volumetric adder ($\tau^b$) and income-graduated fixed charges ($\phi_i$). The objective is to minimize total electricity expenditures while reducing financial inequities across different income brackets. On the follower side, consumers and prosumers (households with DERs) learn their optimal consumption and investment responses under uncertainties in wholesale electricity prices, renewable generation, and demand.

To formally quantify equity in this rate design, we adopt the energy expenditure incidence (EEI) measure~\cite{chen21,chen2023optimaltariff,liu2025equitable}, which represents the share of household income spent on electricity. The EEI for conventional consumers and prosumers at node $i$ is defined as:
\begin{align}
	\mathrm{inc}^{\mathrm{con}}_i &=  \frac{(\text{LMP}_i + \tau^b)d_i + \phi^{\mathrm{con}}_i}{I^{\mathrm{con}}_i}, \label{eq:inc_con}\\
	\mathrm{inc}^{\mathrm{pro}}_i &=  \frac{(\text{LMP}_i + \tau^b)z^b_i + \phi^{\mathrm{pro}}_i + \text{LC}_i}{I^{\mathrm{pro}}_i}, \label{eq:inc_pro}
\end{align}
where $\text{LMP}_i$ is the locational marginal price (the nodal wholesale electricity price), $d_i$ is consumer demand, $z^b_i$ is the prosumer's grid purchase quantity, and $I^{\mathrm{con}}_i$ and $I^{\mathrm{pro}}_i$ represent the average household incomes for each group. The term $\text{LC}_i$ denotes the levelized cost of a prosumer’s solar and storage investment, adjusted to the billing period. (For completeness, prosumers can also sell energy back to the grid at price $\tau^s$ with quantity $z^s_i$).

Finally, for this specific application, we explicitly verify that the model specification—including functional forms, state spaces, and action spaces—satisfies the theoretical conditions required for the existence of a stationary Stackelberg equilibrium and for the convergence of our learning algorithm. This ensures that our learned outcomes correspond to a well-defined equilibrium, distinguishing our approach from purely heuristic or agent-based simulations.

\subsection{Oahu Test System and Model Specification}
This subsection describes the construction of the Oahu test system. We follow the natural building blocks of a power system: first specifying the transmission-level network and utility-scale generation resources (both non-renewable and renewable), then describing residential demand, and finally providing a detailed account of the DERs, including rooftop solar and energy storage. Throughout, we rely on publicly available data whenever possible and introduce simplifying assumptions only where detailed information is unavailable.

\subsubsection{Bulk Transmission Network and Generation Resources}
Due to U.S. critical-infrastructure data protection policies, real transmission topology and line-parameter data are not publicly available. Instead, we use the 37-bus synthetic Oahu, Hawaii system developed in \cite{birchfield}. The dataset provides geographic coordinates for each bus, allowing us to map every utility-scale power plant operated by Hawaiian Electric, the sole utility serving the island, to the nearest bus using plant location data from \cite{powerfacts}. The resulting system includes 26 utility-scale generators: 4 oil-fired units, 2 biomass units, 17 utility-scale solar units (different than the behind-the-meter rooftop PV), and 3 wind units. Because detailed unit-level fuel cost data for Oahu's power plants are not publicly available, we approximate generation costs using standard parameterizations from U.S.~Energy Information Administration (EIA) fuel-cost datasets.\footnote{We use monthly fuel cost and heat-rate data from EIA-923 for oil- and biomass-fired generators across the entire United States for the year 2022. Fuel costs are calculated by combining reported fuel prices with corresponding heat rates, and generator-level quadratic cost functions are estimated via ordinary least squares regression using the Python \texttt{scipy} library. Nationwide data are used instead of only Hawaii in order to mitigate missing and incomplete data entries in regional subsets and to ensure sufficient data coverage for robust parameter estimation. Specifically, for oil and biomass units, we assign quadratic cost functions of the form $C(p) = ap^2 + bp$, where the coefficients are calibrated through regression analysis using EIA fuel cost data. Based on the regression analysis, the base parameters are set to $a = 0.0012\ \$/\mathrm{MW}^2\text{h}$ and $b = 285.51\ \$/\mathrm{MWh}$. To reflect unit-level variability, we perturb the parameters using a triangular distribution $\Delta(0.8, 1.2, 1)$, where $\Delta(\alpha, \beta, m)$ has lower bound $\alpha$, upper bound $\beta$, and mode $m$. Each generator’s cost coefficients are independently sampled from this distribution.}

\begin{figure}[!hbt]
\centering
\includegraphics[width=0.8\linewidth]{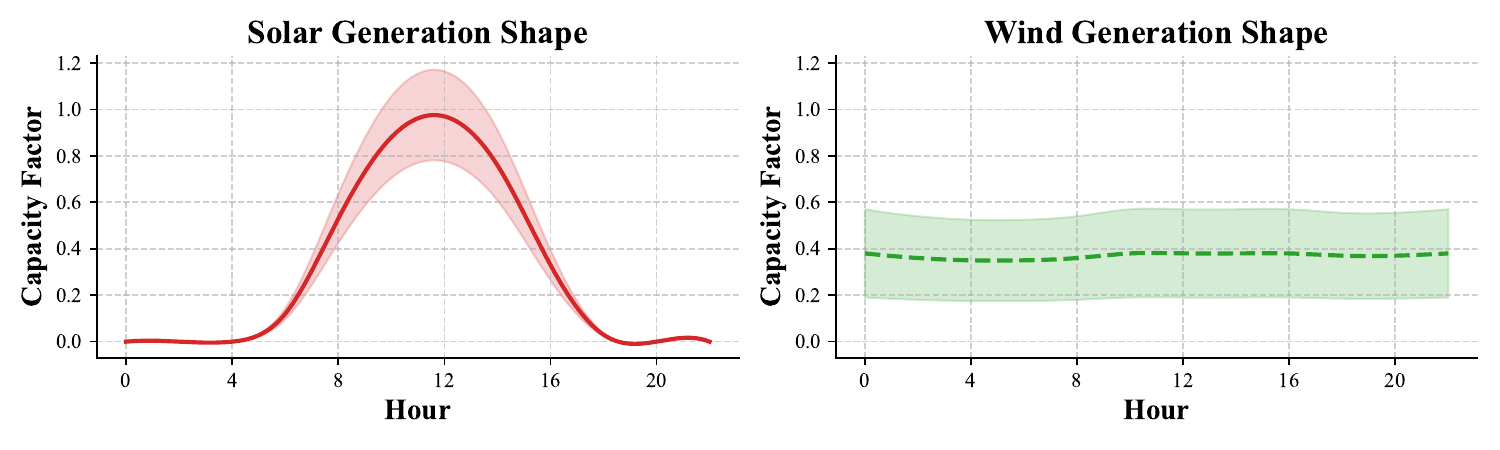}
\caption{Input shapes for solar and wind generators' capacity factors \label{fig:renewable}}
\end{figure}

For solar and wind units, fuel costs are set to zero. Their output variability is captured through capacity factors, which represent the ratio of actual generation to their nameplate capacity over a given period. To model temporal variability without explicitly constructing correlated time series, we use historical hourly capacity-factor profiles as expected values, sourced from utility-scale solar data in \cite{profilesolar} and wind data in \cite{argueso}. At each timestep, the mean capacity factor is perturbed by an independent random noise term: a triangular distribution $\Delta(0.8, 1.2, 1)$ for solar and $\Delta(0.5, 1.5, 1)$ for wind. Figure~\ref{fig:renewable} shows the resulting capacity-factor profiles, with the solid line indicating the mean forecast and the shaded area depicting the uncertainty introduced by the triangular noise.

\subsubsection{Household Demographics and DER Adoption}
We assume that each of the 37 buses in the network hosts two residential agent types: consumers and prosumers. The population of both types is further divided into three income groups, low-, middle-, and high-income, at each bus. The income data from \cite{Census2022ACSST5Y2022.S1901} are reported for the year 2022, while the population and solar installation data obtained from \cite{eia} also correspond to 2022. We therefore choose the base year to be 2022. We classify the low-income group as those households with an annual income less than \$29,999, the middle-income group as those with an annual income from \$30,000 to \$99,999, and the high-income group as those with an annual income of \$100,000 or higher. We then calculate the percentage of households with solar panels, which corresponds exactly to the percentage of prosumers, for each income group based on the 2020 Residential Energy Consumption Survey (RECS) data from \cite{recs20}, which is used to determine the prosumer and consumer population of each income group at each bus in Hawaii. 

Because individual prosumer rooftop PV and storage data are not publicly available, we assign representative system sizes to each income group. Each prosumer is assumed to be equipped with storage capacities of 6, 13, and 20 kWh for low-, middle-, and high-income households, respectively, and corresponding rooftop PV capacities of 5, 10, and 15 kW. These values reflect common residential system sizes: a typical home battery (such as the Tesla Powerwall) provides about 13.5 kWh of usable energy, making our middle tier a direct match; the 6 kWh tier represents smaller single-unit systems marketed for partial backup, and the 20 kWh tier corresponds to multi unit residential storage systems, such as two Powerwall class batteries, commonly adopted in higher consumption homes. The associated PV capacities are based on the assumption that low, middle, and high income households install approximately 10, 20, and 30 panels, respectively, at 500 W per panel \cite{walker_2025}. All batteries are assumed to operate with a one way efficiency of 0.9, which implies a round trip efficiency of $0.9^2 = 0.81$.

\subsubsection{Load Profiles and Demand Uncertainty}

Daily demand profiles are constructed using hourly average data from \cite{coffman2016estimating}, which provides a representative normalized residential load shape for Oahu. In this dataset, hourly values are normalized by the daily peak load; hence, the profile captures the common diurnal pattern of residential consumption. 
We apply this normalized shape uniformly across households and assign different absolute consumption levels to each income group. Specifically, we set reference daily energy levels of 3, 4, and 5 kWh for low-, middle-, and high-income households, respectively. These levels are chosen so that the aggregate load across all agents aligns with Oahu's hourly net load data \cite{{coffman2016estimating}} and reflects the empirical tendency for lower-income households to use less electricity (as documented in the RECS data \cite{recs20}).

\begin{figure}[!hbt]
\centering
\includegraphics[width=0.8\linewidth]{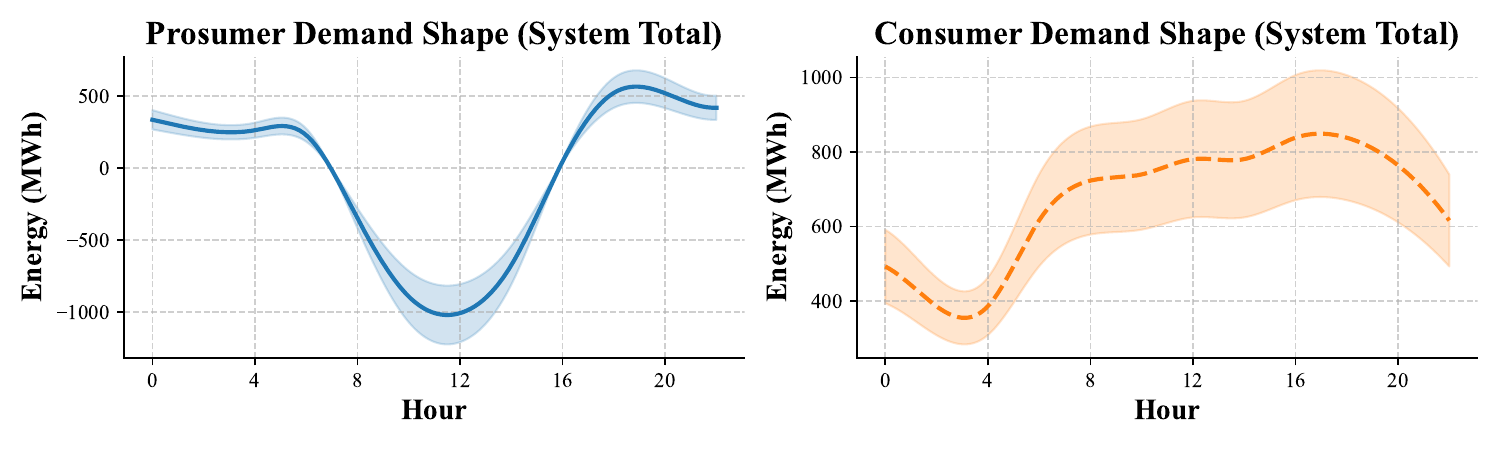}
\caption{Input load demand shapes for both prosumers and consumers. Load demand shapes for both prosumers and consumers (data adapted from \cite{coffman2016estimating}). The shadow areas indicate the noise bound of the shape. \label{fig:demand}}
\end{figure}

To incorporate demand uncertainty, we scale each agent’s average daily demand by a random factor drawn from the triangular distribution $\Delta(0.8, 1.2, 1)$. Figure~\ref{fig:demand} shows the resulting system-level demand profile, where the solid line represents the mean profile and the shaded area depicts the uncertainty created by the triangular scaling.

\subsection{Model Specification} \label{subsec:model}
The complete model is developed based on \cite{he2024evaluating}. The setup follows the framework in Section~\ref{sec:MFE}. The simulation is conducted over a discretized time horizon $t = 1, 2, \ldots$, where each time step corresponds to a two-hour interval. At the beginning of each time step $t$, if the training schedule is triggered, the utility company (leader) is first trained and updates all add-on tariff rates. Subsequently, all aggregators (followers) are trained to learn their battery charging and discharging policies. To adhere to the two-time-scale training, we set the utility company to follow a less frequent training schedule (every 3 days), while aggregators are trained at each time step. After aggregators execute their actions, the resulting aggregate demand bids are submitted to the system operator. At the end of time step $t$, the system operator updates LMPs by solving a simplified economic dispatch (ED) problem over a power network consisting of $N$ buses (substations), $L$ transmission lines, and $G$ fuel-based generators (each generator $g$ has a (strongly convex) quadratic cost function $C_g(\cdot)$ as established earlier). The ED problem is formulated as in~\eqref{eqn:ed-obj}--\eqref{eqn:gen-limit}. The objective at each time $t$ is to find the production for all generators $p_{gt}$ such that total cost is minimized:
\begin{equation}
    \minimize_{\{p_{gt}\}_{g=1}^G} \qquad \sum_{g=1}^G C_g(p_{gt}), \label{eqn:ed-obj}
\end{equation}
and in the meantime satisfying the following constraints for power balance, line flow and generator capacity:
\begin{align}
    & \sum_{g=1}^G p_{gt} = \sum_{n=1}^N D^n_t, \qquad (\mathrm{LMP}^{\text{HUB}}_t), \label{eqn:power-balance} \\
    & -\overline{F}_l \leq \sum_{n=1}^N \mathrm{PTDF}_{ln} \left(\sum_{g \in G_n} p_{gt} - D^n_t\right) \leq \overline{F}_l, (\underline{\phi}_{lt}, \overline{\phi}_{lt}), \; \forall\, l = 1,\ldots,L,
    \label{eqn:flow-limit} \\
    & 0 \le p_{gt} \le \overline{p}_g, \qquad (\underline{\nu}_{gt}, \overline{\nu}_{gt}), \; \forall\, g = 1,\ldots,G, 
    \label{eqn:gen-limit}
\end{align}
where $D^n_t$ denotes the total demand from bus $n$, $\mathrm{PTDF}_{ln}$ represents the power transfer distribution factor associated with line $l$ and bus $n$, and $G_n$ represents the set of generator sets associated with bus $n$. For the associated dual variables, $\mathrm{LMP}^{\text{HUB}}_t$ corresponds to the hub price at time $t$, $\underline{\phi}_{lt}$ and $\overline{\phi}_{lt}$ are the dual variables associated with the lower and upper transmission flow limits of line $l$, respectively, and $\underline{\nu}_{gt}$ and $\overline{\nu}_{gt}$ correspond to the lower and upper generation capacity constraints of generator $g$. The LMP at bus $n \in \{1,\ldots,N\}$ and time $t$, denoted $\mathrm{LMP}^n_t$, is defined as the marginal cost of serving an incremental unit of demand at that bus: $\mathrm{LMP}^n_t$, is derived as follows:
\begin{equation}
    \mathrm{LMP}^n_t := \frac{\partial \mathfrak{L}_t}{\partial D^n_t} = \mathrm{LMP}^{\text{HUB}}_t - \sum_{l=1}^L \mathrm{PTDF}_{ln} (\underline{\mu}_{lt} - \overline{\mu}_{lt}), \label{eqn:lmp}
\end{equation}
where $\mathfrak{L}_t$ represents the Lagrangian function of the ED problem. In establishing the existence and uniqueness of an SS-MFE, a key technical requirement is the Lipschitz continuity of the LMP mapping $\mathrm{LMP}^n_t(\mD_t)$ with respect to the vector of demands 
$\mD_t := (D^1_t, \ldots, D^N_t)$.  Following \cite{feng2025decentralized}, we impose the same linear independence constraint qualification (LICQ) for ED. As shown in \cite{feng2025decentralized}, this condition is sufficient to guarantee that the LMPs are single-valued and Lipschitz continuous of the demand vector $\mD_t$.

\paragraph{MF follower's setup.} Each aggregator (follower) is equipped with finite and discrete state and action spaces. We follow the notation and let $\Actions_F \subseteq [-1, 1]$ denote the action space for each aggregator (and for notation simplicity, we drop aggregator's index), where each action $a_F \in \Actions_F$ represents the proportion of storage capacity to charge (if \( a_F > 0 \)) or discharge (if \( a_F < 0 \)). The state space consists of storage level $e \in [0, 1]$ (as a percentage of capacity), net load $q$ (unrelated to storage charging/discharging), and the current hour of the day $h$. At each time $t$, the state of the follower is defined as the tuple $s_{F,t} := (e_t, q_t, h_t) \in \States_F $ following the state transition kernel $P_F: \States_F \times \Actions_F \mapsto \States_F$. Since the net load is exogenous, and hour of day is deterministic, the only state variable affected by decisions is the storage level, which evolves from time $t$ to $t+1$ according to the following rule after action $a_{F,t}$:
\begin{equation}
    e_{t+1} := \max \{\min \{e_t + a_{F,t}, 1\}, 0 \}.
    \label{eqn:storage-transition}
\end{equation}
Each follower's reward function is defined as the (negative) product of LMP and energy demand, given the storage level $e$, exogenous demand $q$, action $a_F$, and LMP:
\begin{equation}
    r_F(e, q, a_F, \mathrm{LMP}) = - \mathrm{LMP} \cdot \overline{E} \cdot \big( \Phi(e, a_F, \eta) + q \big),
    \label{eqn:rew}
\end{equation}
where \( \eta \in (0, 1] \) is the storage efficiency, $\overline{E}$ is the total battery capacity of the aggregator, and \( \Phi(\cdot) \) adjusts the action for efficiency losses:
\begin{equation}\label{eqn:adjusted-action}
   \Phi(e, a, \eta) = 
   \begin{cases}
   \max \{-e, a \} \cdot \eta, & \text{if } a < 0, \\
   \min \{1 - e, a \} / \eta, & \text{if } a \geq 0.
   \end{cases}
\end{equation}
In sum, for followers, the state and action spaces satisfy the required compactness conditions. The transition kernels are linear and hence continuous. Reward functions are bounded due to finite LMPs and energy demand. While the follower’s reward is linear in price and quantity, the inclusion of an entropy regularization term ensures strong concavity of the value function, satisfying Assumption~\ref{ass:strong-concave-ts}. Moreover, the deterministic MF evolution $\Gamma$ is Lipschitz continuous, as established in \cite{he2024evaluating}, and model parameters are chosen to maintain the contraction condition. 
For any follower, given the MF $\mu_F$ and policy $\pi_F$, $\Gamma$ has the following form:
\begin{equation}
    \Gamma(\mu, \pi) := \zeta \frac{1}{\abs{\States_F}\abs{\Actions_F}} + (1 - \zeta)\sum_{s_F, a_F}\mu_F(s_F, a_F) P_F(\cdot \mid s_F, \cdot) \pi_F(\cdot \mid s_F),
    \label{eq:update-mf}
\end{equation}
where $\zeta \in (0, 1)$ is a the probability of a uniform noise into the MF update as a hyper-parameter. 

\paragraph{Leader's setup.} The leader's state and action spaces are also discretized and hence compact. We define the leader’s reward as the sum of two parts: (i) negative sum of squared EEI differences, and (ii) the welfare. The reward function is defined as follows:
\begin{equation}
    r_L(\tau^b, \tau^s, \{\phi^{\mathrm{pro}}_i\}_{i=1}^N, \{\phi^{\mathrm{con}}_i\}_{i=1}^N) = \sum_{i=1}^N \left(-(\mathrm{inc}^{\mathrm{con}}_i + \mathrm{inc}^{\mathrm{pro}}_i)^2 + \tau^s z^s_i - \tau^b (z^b_i + d_i)\right),
\end{equation}
which is a smooth and concave function. Consequently, the numerical setup satisfies all conditions required by Theorem~\ref{thm:converge-mf-tts}, providing theoretical guarantees for convergence of the two time scale learning procedure to an SS-MFE.

\subsection{Learning-Based Specification} \label{subsec:specification}
In the learning framework, the utility acts as the leader and learns a pricing policy consisting of per-kWh charges for buying and selling electricity, which may differ, together with fixed charges intended to recover T\&D costs. The leader jointly seeks to maximize social surplus, defined as the sum of consumer and producer surplus, and to minimize disparities in EEI across income groups, with EEIs defined by \eqref{eq:inc_con} and \eqref{eq:inc_pro} for consumers and prosumers, respectively.

In principle, fixed charges could be differentiated at the bus level. However, the Oahu network contains 37 buses, and allowing distinct fixed charges for each bus, income group, and consumer type leads to a high dimensional learning problem that is computationally burdensome and exhibits slow convergence in practice. To address this, we aggregate buses into two areas, urban and suburban, based on their geographic and load characteristics. This aggregation reduces the number of fixed charges to be learned to twelve, corresponding to low-, middle-, and high- income consumers and prosumers in urban and suburban areas.  All fixed charges are defined on a daily basis at the area level and are evenly allocated to individual agents. In addition, the utility learns a buy rate and a sell rate that are uniform across all buses, with the buy rate applying to all consumers and prosumers and the sell rate applying to all prosumers, while the two rates may differ.

The leader and each group of followers are trained with Proximal Policy Optimization (PPO)~\cite{schulman2017proximal}. For the training specifications, both leader and followers are given the learning rate 0.0003, discount factor 0.99, entropy coefficient 0.01, batch size 256, number of epochs 10, clip range 0.2, policy network $[24,36]$ and value network $[32,32]$, where the array indicate the number of neurons at each hidden layer of the fully connected neural networks. For the leader and followers, respectively, we set the steps per update to be 360 and 1200, and the training length 3600 and 2400. Each simulated day consists of 12 time steps, corresponding to 2-hour intervals. We assume the utility updates its policy on a daily basis, that is, every 12 time steps, while aggregators update their policies at every time step. Each simulation runs for 50 days and is repeated five times with different random seeds. All experiments were conducted on Windows 11 with a 13th Gen Intel Core i7 13700KF processor with 24 cores and an NVIDIA GeForce RTX 4070 GPU.

\subsection{Numerical Results}
\subsubsection{Validation with Benchmark Results}
To evaluate the performance of the learned tariffs, we define a baseline case based on the current retail rates used by Hawaiian Electric. Specifically, we adopt the \textit{Shift and Save Schedule R}~\cite{scheduler} together with the \textit{Smart Renewable Energy} program~\cite{sre}. Under Schedule R, retail electricity prices are set at 15.75, 31.51, and 47.26 \textcent/kWh during the daytime period from 9 AM to 5 PM, the overnight period from 9 PM to 9 AM, and the evening peak period from 5 PM to 9 PM, respectively. Compensation for electricity sold back to the grid by prosumers under the Smart Renewable Energy program is 13.5, 18.9, and 32.9 \textcent/kWh for the same time periods. Together, these prices constitute a time-of-use (TOU) tariff. Schedule R also includes a fixed monthly charge of \$16.32, consisting of a \$6.94 customer charge and a \$9.38 grid access charge. All rates are reported prior to adjustment for inflation using the consumer price index.

In this TOU benchmark, the tariff structure is taken as given and no leader optimization is performed, while consumers and prosumers still learn their optimal consumption and storage policies. In contrast, our learning based scenario employs real time prices for follower decisions and endogenously optimizes the leader’s volumetric and income graduated fixed charges. Comparing these two regimes therefore captures both the behavioral effects of real time pricing relative to time of use pricing and the additional gains arising from learning the tariff structure itself.

To be consistent with our modeling framework, we further decompose the existing retail rates into energy cost and non-energy components. Hawaiian Electric provides a detailed breakdown of residential bill components for Oahu customers~\cite{knowurrates}. Among these components, the \textit{Energy Cost Recovery (ECR)} charge reflects fuel related energy costs and serves as a proxy for the underlying wholesale electricity price, while the remaining components recover T\&D costs. The ECR rate is publicly filed each month and varies over time. According to the latest filing~\cite{ecrf}, the ECR for Oahu in October 2025 is 17.782 \textcent/kWh for residential customers, which we approximate as 20 \textcent/kWh for simplicity.

Using this approximation, we define the \textit{buy rate} as the portion of the retail purchase price exceeding the ECR, given by $\max(\text{retail buy rate} - \text{ECR}, 0)$. Under Schedule R, this yields buy rates of 0, 11.51, and 27.26 \textcent/kWh for the daytime, overnight, and evening peak periods, respectively. Similarly, we define the \textit{sell rate} as the difference between the retail selling price and the ECR, given by $\min(\text{retail sell rate} - \text{ECR}, 0)$, resulting in sell rates of -6.5, -1.1, and 0 \textcent/kWh for the same periods. The negative sell rates reflect a regulatory design choice under which electricity sold by prosumers is compensated below the average energy cost during periods of low prices or low demand, in order to address equity concerns and account for continued reliance on T\&D infrastructure.

The baseline case is finalized by fixing these buy and sell rates and training only the followers to learn their charging and discharging policies. After CPI adjustment to real 2022 USD, the resulting T\&D cost is approximately \$1.27 million per day, calculated as the total daily utility revenue under the specified rates. This corresponds to roughly \$464 million annually, which is comparable in magnitude to the roughly \$800 million in annual transmission, distribution, customer, and administrative costs reported statewide across Hawaiian Electric service territories in \cite{budget}. This statewide figure includes multiple islands beyond Oahu, which alone accounts for approximately 60 to 70 percent of total electricity demand, and also covers commercial and industrial customers, whereas our estimate reflects only the residential sector on Oahu. Taken together, these differences indicate that the \$464 million annual estimate is well within the expected range. This value then enters the learning-based formulation as an explicit constraint, ensuring that the utility’s tariff revenue, net of energy costs, is at least \$1.27 million per day, or equivalently \$464 million annually. 

According to Hawaiian Electric’s \emph{Adequacy of Supply Report} \footnote{Hawaiian Electric Company, \emph{Adequacy of Supply Report}, filed with the Hawaii Public Utilities Commission, 2023. Available at \url{https://puc.hawaii.gov/wp-content/uploads/2023/05/Adequacy-of-Supply-HECO-2023.pdf}.}, in 2022, the system peak on Oahu occurred on Wednesday, October 10, 2022, at approximately 5:52 p.m. and reached 1,074 MW (net load). This value is very close to the peak demand observed in our benchmark simulation. While the actual system peak includes commercial, industrial, and agricultural loads, not just residential consumption, our simulated results remain well within a realistic range and capture the correct order of magnitude for Oahu’s peak system demand.

\subsubsection{Follower's Learning Result}\label{subsec:numerical-follower}
\begin{figure}[!hbt]
\centering
\includegraphics[width=0.8\linewidth]{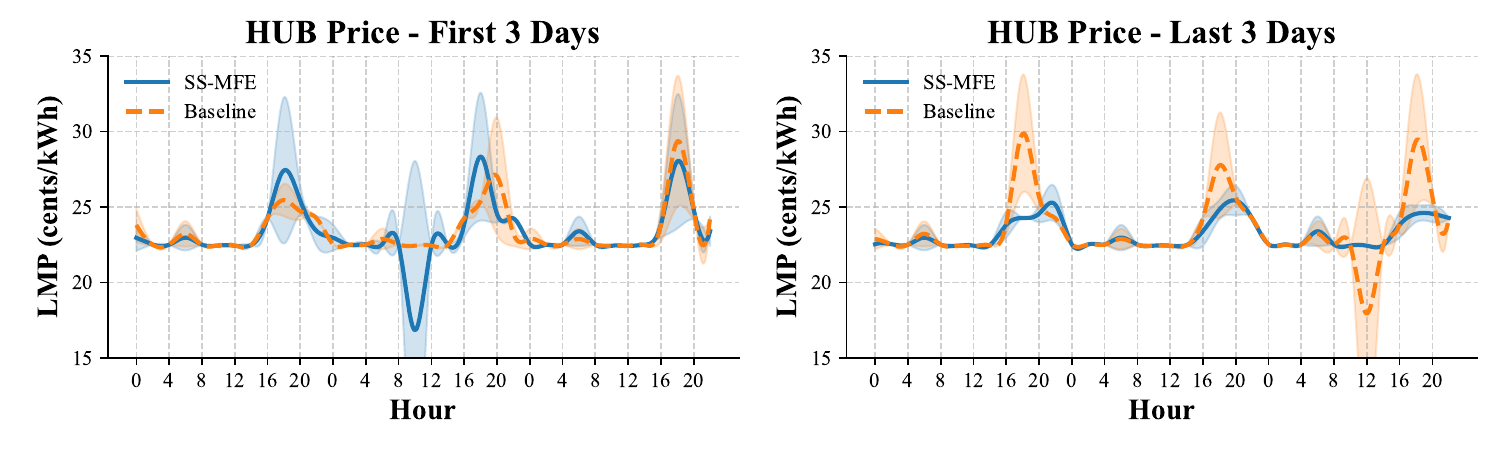}
\caption{Comparison of HUB prices in the SS-MFE learning and baseline case during the first 3 days (left) and the last 3 days (right). RL reduces price volatility and leads to more stable daily patterns. \label{fig:lmp}}
\end{figure}

In this section, we compare the followers' results learned under two scenarios: SS-MFE and baseline. The first comparison, as shown in Figure~\ref{fig:lmp}, shows the wholesale electricity prices (\emph{i.e.}, LMP) at the Hub location. The Hub is a collection of locations intended to represent an uncongested price for electricity. Under SS-MFE, volatility reduces significantly, and daily patterns stabilize. 

\begin{figure}[!hbt]
\centering
\includegraphics[width=0.8\linewidth]{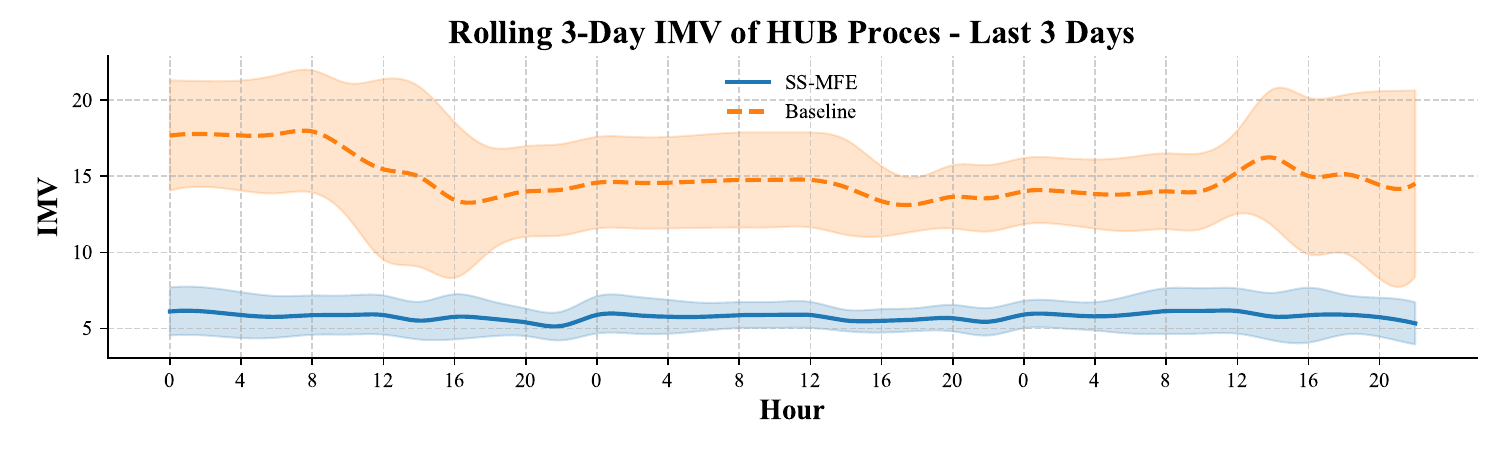}
\caption{Comparison of IMV of the last 3 days between the SS-MFE learning and baseline case. Shadow areas show the 1-sigma error bounds across all simulations. \label{fig:imv}}
\end{figure}

To better measure the impacts of energy storage coupled with the RL algorithms on LMP volatility, we adopt \textit{incremental mean volatility} (IMV) from \cite{roozbehani} as the metric. For a sequence of LMPs $\{\text{LMP}_t\}_{t=1}^{\infty}$, the IMV is defined as
\begin{equation} \label{eqn:measure}
  \text{IMV} = \lim_{T \to \infty} \frac{1}{T} \sum_{t=1}^T \Big| \text{LMP}_{t+1} - \text{LMP}_{t} \Big|.
\end{equation}
Figure \ref{fig:imv} shows the IMV of the last 3 days between the two scenarios. Results indicate that SS-MFE achieves a significant reduction in IMV, approximately 10 units lower on average, which shows notably less volatile and more stable electricity prices.

\begin{figure}[!hbt]
\centering
\includegraphics[width=0.8\linewidth]{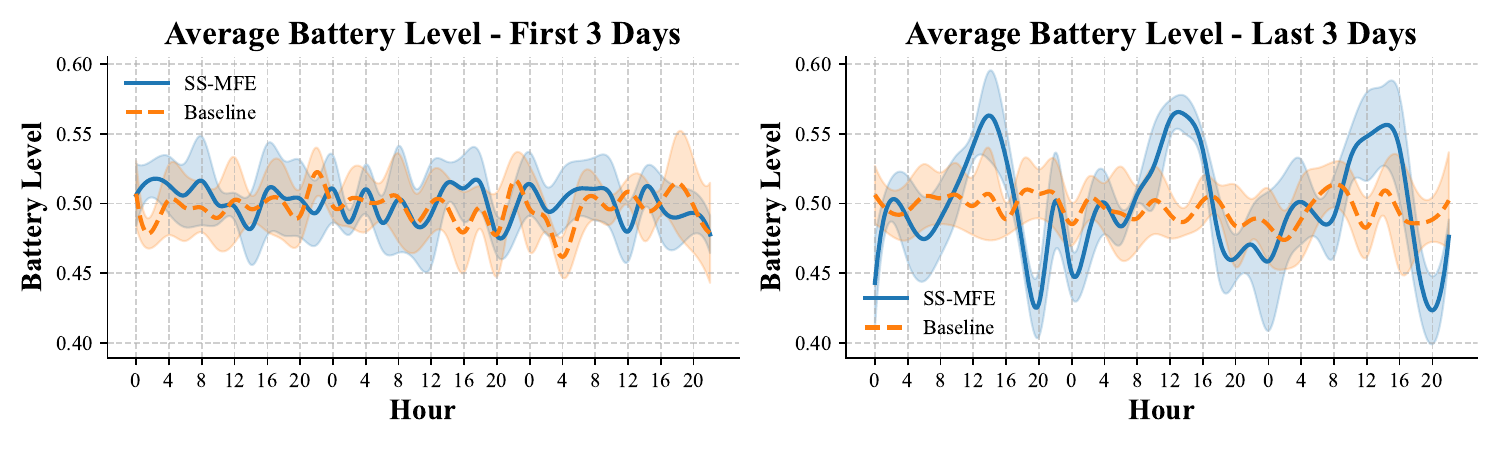}
\caption{Comparison of average battery level across the system of the first and last 3 days of learning. Shadow areas show the 1-sigma error bounds across all simulations. \label{fig:battery}}
\end{figure}

Figure~\ref{fig:battery} shows the average battery level across the system of the first and last 3 days of learning. We observe that the aggregators' charging/discharging actions has a similar patterns at the end of SS-MFE learning, which indicates the convergence of followers' actions. Comparing to the baseline case, at SS-MFE, prosumers are more active in charging/discharging their batteries. 

\begin{figure}[!hbt]
\centering
\includegraphics[width=0.8\linewidth]{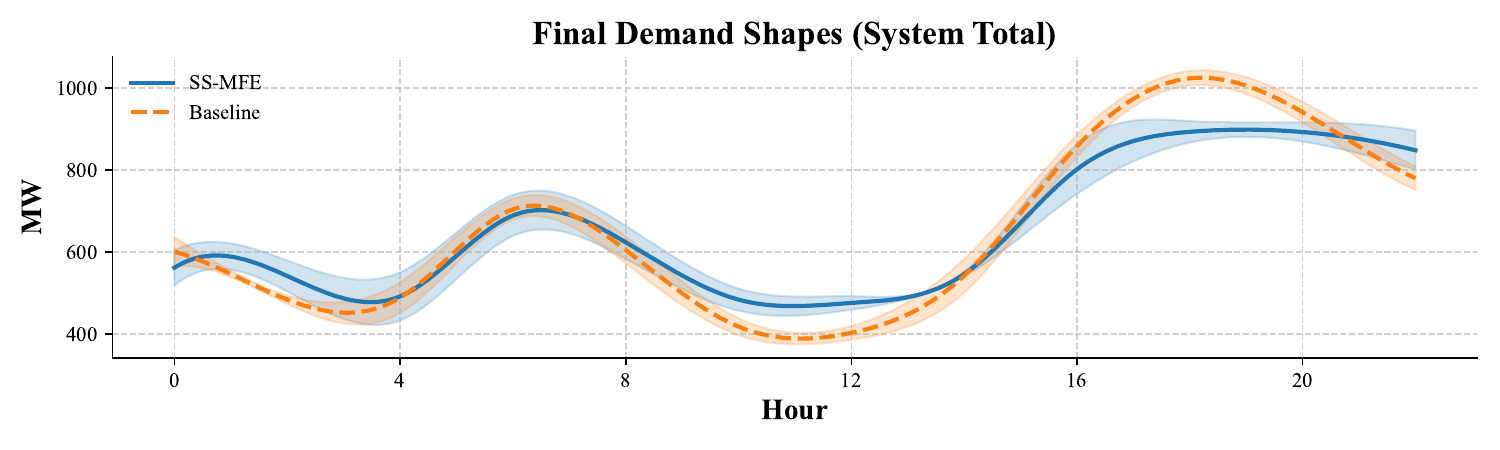}
\caption{Comparison of daily demand shapes between SS-MFE learning and baseline case. The vertical axis indicates the total instantaneous demand (in MW) across the system population. The shaded areas indicate one standard deviation error bounds computed over the last 3 days upon convergence and all simulation runs. \label{fig:final-shape}}
\end{figure}

To further understand how SS-MFE influences consumption behavior, we examine the resulting net demand profiles and compare them to the baseline case. As shown in Figure~\ref{fig:final-shape}, we observe that the SS-MFE case reduced the ramping needs of about 200 MW: in SS-MFE, the daily peak and valley difference is approximately 400 MW, while in the baseline case, the difference is around 600 MW. The reduction from SS-MFE is approximately 200 MW, which is approximately 1/6 of the total generation capacity (1200 MW) of Hawaiian Electric. 

\subsubsection{Leader's Learning Result}\label{subsec:numerical-leader}
\begin{figure}[!hbt]
\centering
\includegraphics[width=0.8\linewidth]{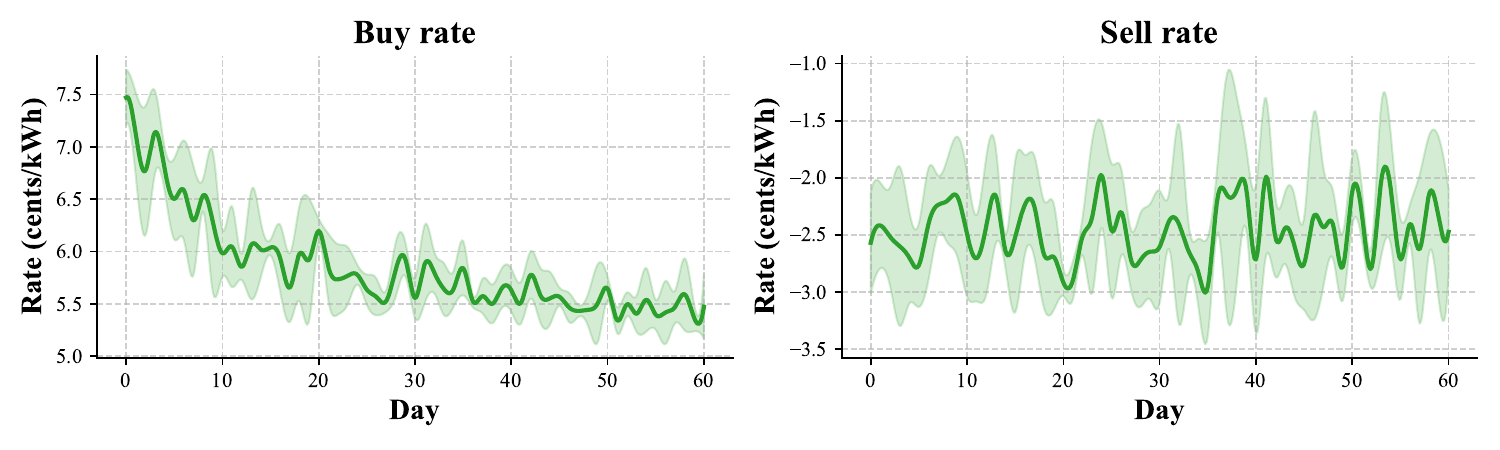}
\caption{Learned variable charges for buy (left) and sell (right) rates. The price converges through learning especially for buy rates. Sell rates oscillates between -2 and -3 due to the fact the selling occurs much less frequent and that the algorithm starts with a value close to convergence. Shaded regions represent one standard deviation across simulation runs. \label{fig:buy-sell}}
\end{figure}

\begin{figure}[!hbt]
\centering
\includegraphics[width=0.8\linewidth]{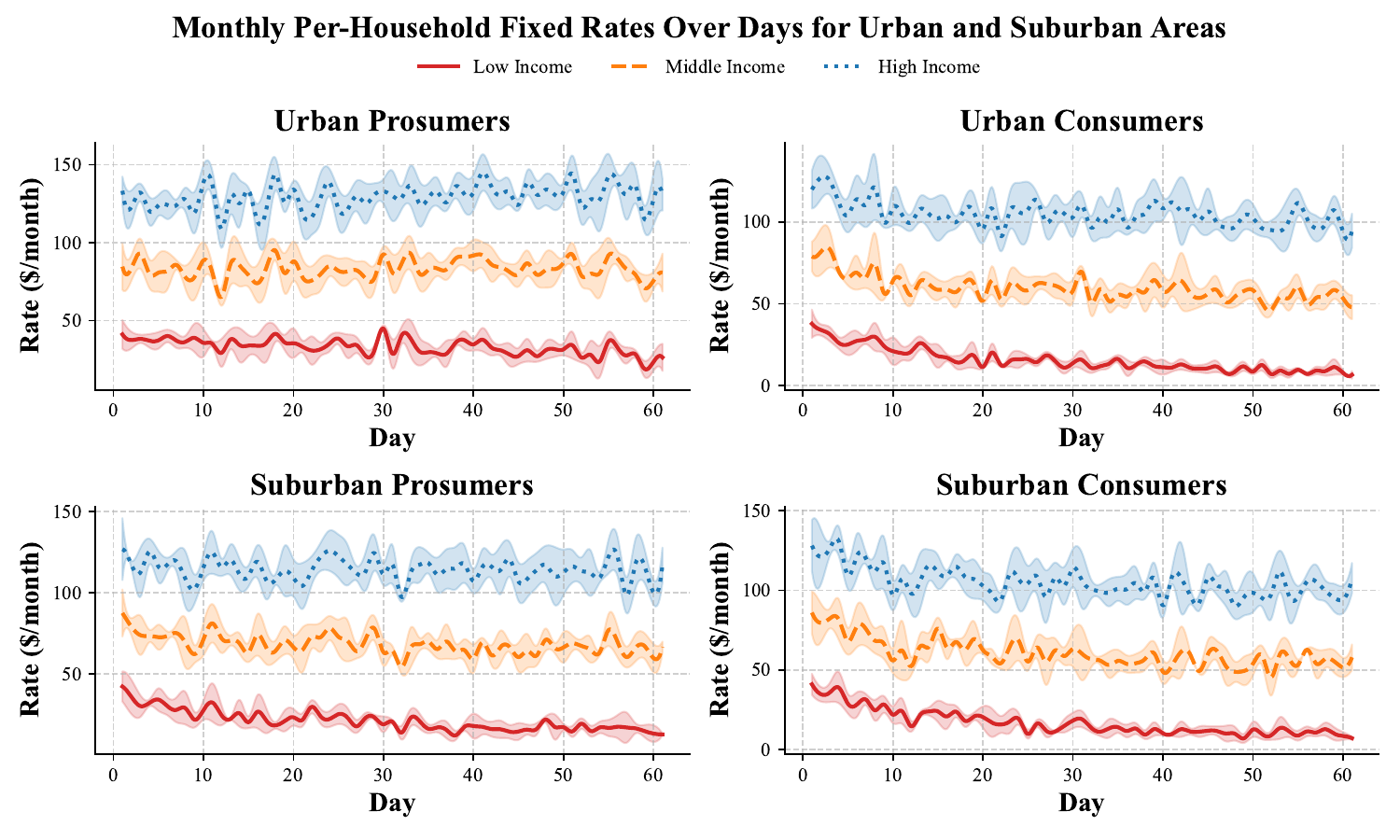}
\caption{Learned daily fixed charges for urban (left) and suburban (right) areas. Shaded regions represent one standard deviation across simulation runs. \label{fig:fixed-rates}}
\end{figure}

Figure~\ref{fig:buy-sell} shows the learned per-MWh variable rates that are adders on top of the wholesale electricity prices (LMPs) over the course of training. Figure~\ref{fig:fixed-rates} shows the learned electricity daily fixed charges over training. Over time, the utility's policy converges to a pricing structure in which prosumers are charged more than consumers, helping to align payment responsibility with ability to pay and maintain energy equity. However, in our experiment, the buying rate adder's curve shows a more obvious convergence trend, but the selling rate curve oscillates in between the value -2 to -3. This is due to the fact that from the aggregator's perspective, selling to the grid occurs much less than buying, and that our algorithm started with a value that is potentially close to the convergence. This result aligns with our intuition that prosumers have the advantage of utilizing the LMP information to further reduce their electricity charge; lowering the buy rates and setting a lower fixed rates for consumers can help consumers reduce charge and minimize the EEI difference.

\begin{figure}[!hbt]
\centering
\includegraphics[width=0.8\linewidth]{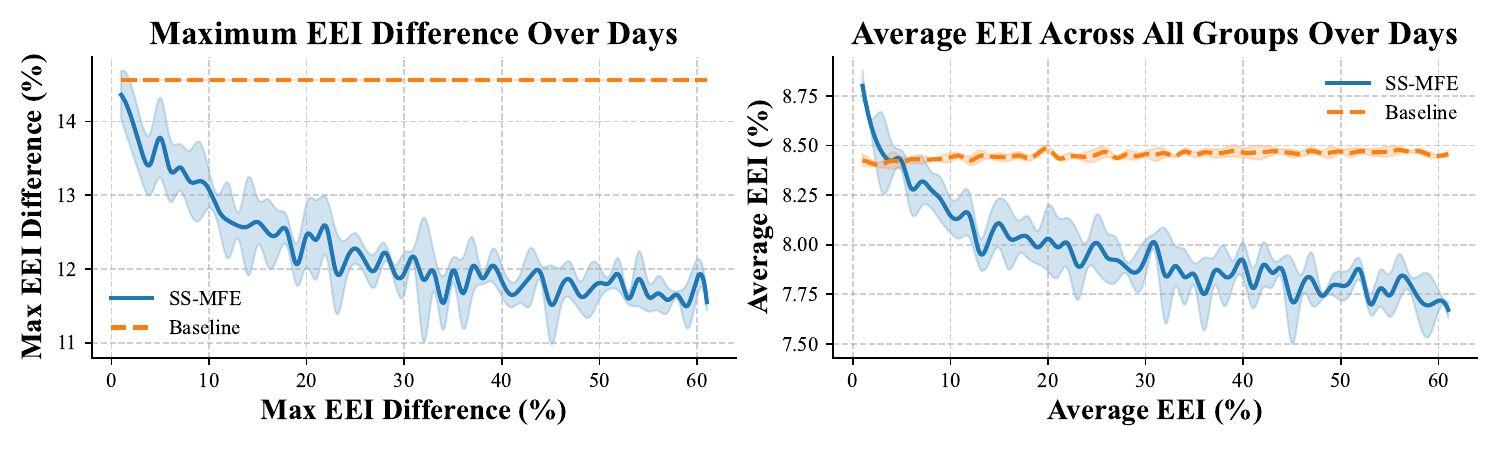}
\caption{Maximum daily EEI difference over time across all 3 income groups and prosumer/consumer types in both urban and suburban areas. The learned policy reduces the maximum EEI difference, indicating improved income-based equity. Shaded regions represent one standard deviation across all simulation runs.\label{fig:eei}}
\end{figure}

We now show the maximum EEI difference through cross comparison across all 3 income groups and prosumer/consumer types in both urban and suburban areas, and the average EEI of all groups in Figure~\ref{fig:eei}. Under our experimental setup, the utility company's optimal strategy from SS-MFE learning reduces the maximum EEI difference from approximately 15\% to 12\%, while the baseline case is 15\%. The average EEI in SS-MFE also decreases from 8.75\% to 7.75\%, while the baseline case maintains a constant of 8.50\%. This shows that SS-MFE learning can improve equity across different income groups and customer types. We note that the EEI values are typically small since energy spending constitutes only a minor portion of total household income \cite{recs20}.

We compare each group's EEI between the SS-MFE learning framework and the baseline case. The reported values are averaged over the last 10 days across all simulations where the EEIs converge. As shown in Table~\ref{tab:cost}, the SS-MFE learning induces clear distributional changes in EEI between consumers and prosumers. Compared to the baseline, the learned SS-MFE rates reduce EEI for low-income prosumers and consumers in both urban and suburban areas, with the largest reductions observed among low-income consumers. For middle-income households, EEI changes are modest: prosumers experience a slight increase, while consumers see a small decrease. Finally, EEI increases across all high-income groups, indicating higher costs for wealthier agents who typically contribute more to peak demand or grid stress as well as larger DER equipment. These patterns suggest that the learned SS-MFE pricing scheme internalizes system-level costs associated with DER generation, encouraging prosumers to better coordinate their production and storage with grid conditions. Consumers, in turn, benefit from lower and more stable rates, enhancing affordability and equity. Overall, the SS-MFE learning framework yields a more balanced and efficient allocation of electricity costs, shifting part of the economic burden from consumers to prosumers in a manner that promotes both fairness and grid stability.

\begin{table}[!hbt]
    \centering
    \small
    \caption{Comparison between SS-MFE and Baseline Monthly EEI}
    \label{tab:cost}
    \begin{subtable}[t]{\textwidth}
        \centering
        \begin{tabular}{l rrr rrr}
            \toprule
            & \multicolumn{3}{c}{\textbf{Urban Prosumer}} & \multicolumn{3}{c}{\textbf{Urban Consumer}} \\
            \cmidrule(lr){2-4} \cmidrule(lr){5-7}
            \textbf{Metric} & \textbf{Low} & \textbf{Middle} & \textbf{High} & \textbf{Low} & \textbf{Middle} & \textbf{High} \\
            \midrule
            \textbf{SS-MFE (\%)}     & 12.23 & 7.02 & 4.11 & 14.65 & 6.13 & 3.12 \\
            \textbf{Baseline (\%)}   & 13.01 & 6.96 & 3.96 & 17.62 & 6.44 & 3.05 \\
            \textbf{Difference (\%)} & -0.78 & 0.06 & 0.15 & -2.97 & -0.31 & 0.07 \\
            \bottomrule
        \end{tabular}
    \end{subtable}
    
    \vspace{1em} 
    
    \begin{subtable}[t]{\textwidth}
        \centering
        \begin{tabular}{l rrr rrr}
            \toprule
            & \multicolumn{3}{c}{\textbf{Suburban Prosumer}} & \multicolumn{3}{c}{\textbf{Suburban Consumer}} \\
            \cmidrule(lr){2-4} \cmidrule(lr){5-7}
            \textbf{Metric} & \textbf{Low} & \textbf{Middle} & \textbf{High} & \textbf{Low} & \textbf{Middle} & \textbf{High} \\
            \midrule
            \textbf{SS-MFE (\%)}     & 11.30 &  6.65 & 4.02 & 14.67 &  6.09 & 3.15 \\
            \textbf{Baseline (\%)}   & 12.90 &  6.86 & 3.97 & 17.42 &  6.34 & 3.08 \\
            \textbf{Difference (\%)} & -1.60 & -0.21 & 0.05 & -2.75 & -0.25 & 0.07 \\
            \bottomrule
        \end{tabular}
    \end{subtable}
    
    \vspace{1ex}
    \raggedright
    \small{\emph{Note: } SS-MFE EEI uses results from our training while baseline EEI is computed with real-world rates set by Hawaiian Electric. The last row measures the difference = SS-MFE EEI - baseline EEI. Table reports the average value across 5 simulations.}
\end{table}

\begin{figure}[!hbt]
\centering
\includegraphics[width=0.8\linewidth]{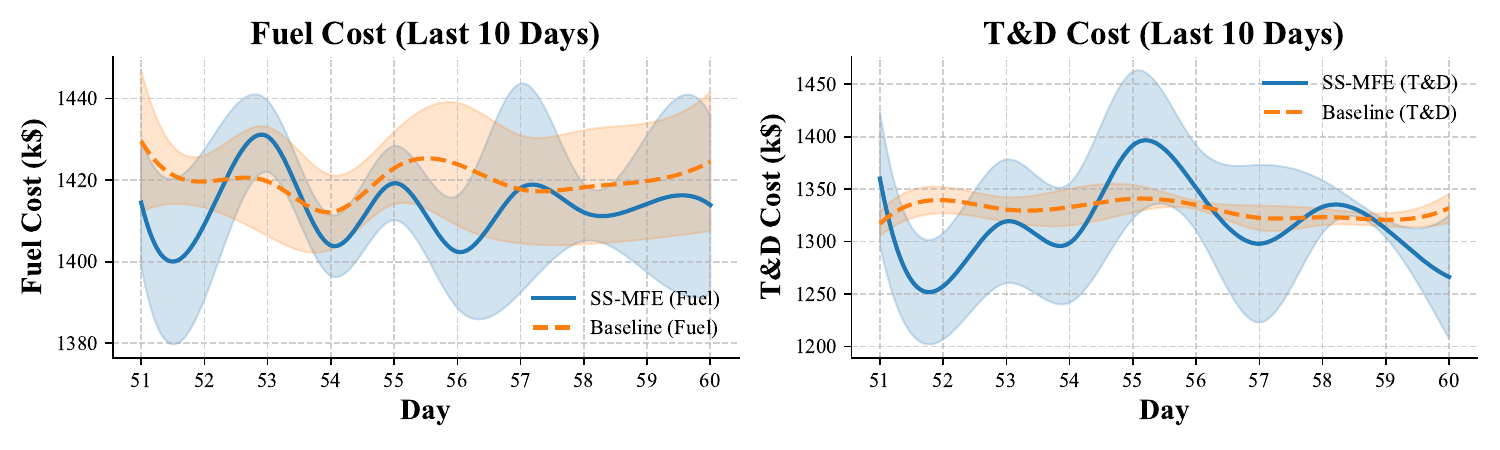}
\caption{Learned SS-MFE cost uses results from our training while baseline cost is based on learning result with preset real-world rates set by Hawaiian Electric. Shaded regions represent one standard deviation across all simulation runs.\label{fig:cost}}
\end{figure}

Finally, Figure~\ref{fig:cost} compares daily grid-level fuel generation costs between the learned SS-MFE and the baseline case over the last 10 days of simulation. Overall, the results indicate that, on average (across simulations), SS-MFE achieves slightly lower total fuel costs and T\&D costs, reflecting improved operational efficiency and coordination between distributed agents and the grid. Specifically, the SS-MFE framework reduces daily fuel costs by roughly \$5-20k compared to the baseline on most days, indicating that the learned equilibrium effectively minimizes reliance on fuel units. T\&D costs also show a reduction in the total number.

In sum, these findings show that the SS-MFE learning not only yields socially fairer pricing outcomes (as seen in the EEI results) but also delivers system-level efficiency gains. By jointly optimizing leader-follower interactions, the framework outputs smoother LMP patterns, lowers overall fuel use, and balances network costs. It shows the potential of Stackelberg-MF learning for sustainable power system management.
\section{Conclusion} \label{sec:conclusion}
This paper establishes the existence and uniqueness of stationary Stackelberg equilibria under standard continuity and Lipschitz conditions. We introduce a learning algorithm for equilibrium computation, and to analyze its convergence, we incorporate Boltzmann policy smoothing and reward regularization. These techniques allow us to prove that the learning dynamics converge to a unique fixed point. We then extend the framework to settings with a MF population of followers, resulting in a tractable formulation of SS-MFE. This enables applications to large-scale policy design problems, such as those modeled in the AI Economist framework. As a proof of concept, we apply our method to the design of retail electricity tariffs in a stylized power system, showing that the learned policy achieves favorable trade-offs among economic efficiency and equity across customer classes. An important direction for future work is to extend the current framework to include heterogeneous MF followers, where agents may differ in preferences, constraints, or dynamics. Another major extension is to have a finite number of fully strategic followers. In such cases, the lower level of the game requires computing a Nash equilibrium among the followers in response to the leader's policy, which introduces significant theoretical and algorithmic challenges. 

\bibliographystyle{IEEEtran} 
\bibliography{references}

\appendix
\section{Appendix}
\subsection{Contraction Mapping and Fixed-Point Theorems} \label{appendix:fixed-point-proof}
Our proof relies on the well-known Banach fixed point theorem, for which we first restate the definition of a contraction mapping, and then the Banach fixed point theorem.
\begin{definition}[Contraction Mapping]
    Let $(\mathcal{X}, d)$ be a non-empty complete metric space, where $d$ is a metric on $\mathcal{X}$. A map $T: \mathcal{X} \mapsto \mathcal{X}$ is called a contraction mapping on $\mathcal{X}$ if for any $x, y \in \mathcal{X}$, there exists a constant $c \in [0, 1)$ such that $d(T(x), T(y)) \leq cd(x, y)$.
\end{definition}
\begin{theorem}[Banach Fixed-Point Theorem~\cite{Banach1922}] \label{thm:banach}
Let $(\mathcal{X}, d)$ be a non-empty complete metric space, and let $T: \mathcal{X} \to \mathcal{X}$ be a contraction mapping. Then $T$ admits a unique fixed point $x^* \in \mathcal{X}$ such that $T(x^*) = x^*$.
\end{theorem}
In our work, we choose the distance function $d$ to be the $\ell_1$-norm.

\subsection{Properties of Shannon Entropy and Softmax Functions} \label{appendix:softmax}
\begin{lemma}[Bounds on Shannon Entropy on a Finite Support]\label{lem:shannon-bounds}
Let $\mathcal{X}$ be a finite set with $|\mathcal{X}|=m\ge 1$, and let $p$ be any probability distribution on $\mathcal{X}$. Define the Shannon entropy $H(p):= -\sum_{x\in\mathcal{X}} p(x)\log p(x)$ with the convention $0\log 0 := 0$. Then $0 \le H(p) \le \log m$.
\end{lemma}

\begin{proof}
For any $x\in\mathcal{X}$ with $p(x)>0$, we have $\log p(x)\le 0$, hence $-p(x)\log p(x)\ge 0$. With the convention $0\log 0:=0$, every summand is nonnegative, so $H(p)\ge 0$. Next, consider the Kullback--Leibler divergence from $p$ to the uniform distribution $u$ on $\mathcal{X}$, where $u(x)=1/m$:
\begin{equation*}
\mathrm{KL}(p\|u) :=\sum_{x\in\mathcal{X}} p(x)\log\frac{p(x)}{u(x)} =\sum_{x\in\mathcal{X}} p(x)\log\big(p(x)m\big) =\sum_{x\in\mathcal{X}} p(x)\log p(x) + \log m.
\end{equation*}
Rearranging gives $\mathrm{KL}(p\|u)=\log m - H(p)$. As $\mathrm{KL}(p\|u)\ge 0$ for any $p$, it follows that $H(p)\le \log m$.
\end{proof}

The first lemma is to show the Lipschitz continuity of softmax under $\ell_1$-norm.
\begin{lemma}[Lipschitz Continuity of Softmax] \label{lemma:lip-softmax}
Let $\softmax_c: \RR^n \mapsto [0, 1]^n$ be the softmax function with temperature $c > 0$ defined by $\softmax_c(\vx)_i = \frac{e^{c x_i}}{\sum_{j=1}^n e^{c x_j}}$. Then $\softmax_c$ is $\sqrt{n} c$-Lipschitz continuous with respect to the $\ell_1$ norm; that is, $\normof{\softmax_c(\vx) - \softmax_c(\vy)}_1 \leq \sqrt{n} c \normof{\vx - \vy}_1$ for all $\vx, \vy \in \RR^n$.
\end{lemma}
\begin{proof}
    Choose arbitrary vectors $\vx, \vy \in \RR^n$. We first show that the $\softmax_c$ function is $c$-Lipschitz with respect to $\ell_2$-norm. For brevity, in this proof only, we let $s_i = \softmax_c(\vx)_i$. We start by computing the Jacobian matrix of $\softmax(\vx)$, denoted as $J_c(\vx)$, whose entries are given by:
    \begin{equation*}
        J_c(\vx)_{ij} = \frac{\partial s_i}{\partial x_j} = \begin{cases}
            c s_i(1 - s_i), & \text{if } i=j, \\
            -c s_i s_j, & \text{if } i\neq j,
        \end{cases}
    \end{equation*}
    and for any vector $\vu \in \RR^n$, we have, $\vu^\top J_c(\vx) \vu = c \left( \sum_{i=1}^n s_i u_i^2 - \left(\sum_{i=1}^n s_i u_i\right)^2 \right)$. By applying Jensen's inequality to the convex function $x \mapsto x^2$ and by viewing $\softmax$ as a probability distribution, we have $\left(\sum_{i=1}^n s_i u_i\right)^2 \leq \sum_{i=1}^n s_i u_i^2$. This implies that $0 \leq \vu^T J_c(\vx) \vu \leq c \sum_{i=1}^n s_i u_i^2 \leq c \normof{\vu}_2$. The eigenvalues of $J_c(\vx)$ lie between 0 and $c$. Hence, its spectral norm satisfies that $\normof{J_c(\vx)}_2 \leq c$. By the Mean Value Theorem for vector-valued functions, we have $\normof{\softmax_c(\vx) - \softmax_c(\vy)}_2 \leq \sup_{t \in [0, 1]} \normof{J_c(\vy + t(\vx - \vy))}_2 \cdot \normof{\vx - \vy}_2 \leq c\normof{\vx-\vy}_2$. Since $\normof{\vx}_2 \leq \normof{\vx}_1 \leq \sqrt{n} \normof{\vx}_2$, we have $\normof{\softmax_c(\vx) - \softmax_c(\vy)}_1 \leq \sqrt{n} \normof{\softmax_c(\vx) - \softmax_c(\vy)}_2 \leq \sqrt{n} c \normof{\vx - \vy}_1$.
\end{proof}

We now show the closeness of softmax distribution and the uniform argmax distribution.
\begin{lemma}[Softmax–Argmax Closeness]\label{lemma:softmax-argmax-u}
    Let $\vx \in \RR^n$. Then for any $c > 0$, the $\ell_1$ distance between $\softmax_c(\vx)$ and $\argmax(\vx)$ is bounded as $\normof{ \softmax_c(\vx) - \argmax(\vx) }_1 \leq 2n e^{-c \delta}$, where we define $\delta := \min_{j : x^j < x^*} (x^* - x_j) > 0$ as the minimum gap between the top value $x^*$ and the next largest value, and $\delta := \infty$ when all values are equal in $\vx$.
\end{lemma}
\begin{proof}
    For any $\vx \in \RR^n$, let the $\argmax(\vx) = \{ \ones_k \}_{ k \in \mathcal{K} }$, where $\mathcal{K} = \{k : x_k = \max_{i = 1, \cdots, n} x_i\}$ is the index set of the largest entries of $\vx$, and $\ones_k \in \RR^n$ be the vector where the $k$-th entry equals one and all else equal zero. Then for any $k \in \mathcal{K}$, $\ones_k$ can be viewed as a deterministic distribution over $\vx$ by following any tie-breaking rule with which a unique maximal entry is selected. For each $i = 1, \ldots, n$, the softmax function with temperature $c > 0$ is defined as $\softmax_c(\vx)_i = \frac{e^{c \vx_i}}{\sum_{j=1}^n e^{c x_j}}$. For simplicity, let $Z = \sum_{j=1}^n e^{c x_j} = e^{c x_k} + \sum_{j: j \neq K} e^{c x_j}$. For any $k \in \mathcal{K}$, we have $\normof{ \softmax_c(\vx) - \ones_k }_1 = \left( 1 - \frac{e^{c x_k}}{Z} \right) + \frac{\sum_{j: j \neq K} e^{c x_j}}{Z} = \frac{2\sum_{j: j \neq K} e^{c x_j}}{Z}$. Now let $\delta = \min_{j : j \neq K} (x_k - x_j) > 0$. Then for all $j \neq K$, $x_j \le x_k - \delta$, we have $e^{c x_j} \le e^{c (x_k - \delta)} = e^{-c \delta} e^{c x_k}$. As a result, the numerator satisfies that $\sum_{j : j \neq K} e^{c x_j} \le (n - 1) e^{-c \delta} e^{c x_k} \leq n e^{-c \delta} e^{c x_k} $, and the denominator satisfies that $Z = e^{c x_k} + \sum_{j : j \neq K} e^{c x_j} \ge e^{c x_k}$. Then we get $\normof{ \softmax_c(\vx) - \ones_k }_1 \leq \frac{2n e^{-c \delta} e^{c x_k}}{e^{c x_k}} \le 2n e^{-c \delta}$, which concludes the proof.
\end{proof}

\end{document}